\documentclass{aa}  
\usepackage{graphicx}
\usepackage{gensymb}
\usepackage{txfonts}
\usepackage{fixltx2e}
\usepackage{xcolor}
\usepackage{float}
\usepackage{caption}

\begin{document}

   \title{Ionized regions in the central arcsecond of NGC 1068 \protect\footnotemark }

   \subtitle{YJHK spatially resolved spectroscopy}

   \author{
          P. Vermot
          \inst{1} 
          \and 
          B. Barna
          \inst{2}
          \and
          S. Ehlerov\' a
          \inst{1}
          \and 
          M. R. Morris
          \inst{3}
          \and
          J. Palou\v s
          \inst{1}
          \and
          R. W\" unsch
          \inst{1}
          }
     \institute{Astronomical Institute of the Czech Academy of Sciences,
              Bo\v{c}n\'{\i} II 1401/1, 140 00 Prague\\
              \email{pierre.vermot@asu.cas.cz}
              \and
           Physics Institute, University of Szeged, D\'{o}m t\'{e}r 9, Szeged, 6723, Hungary
              \and
           Department of Physics and Astronomy, University of California, Los Angeles, CA 90095-1547, USA
             }

   \date{Received September 15, 1996; accepted March 16, 1997}

 
  \abstract
   {Several bright emission line regions have been observed in the central 100 parsecs of the active galaxy NGC 1068.}
   {We aim to determine the properties and ionization mechanism of three regions of NGC 1068: the nucleus (B) and two clouds located at 0.3" and 0.7" north of it (C and D).}
   {We combined SPHERE (0.95 - 1.65 $\mu m$) and SINFONI (1.5 - 2.45 $\mu m$) spectra for the three regions B, C, and D. We compared these spectra to several CLOUDY photoionization models and to the MAPPINGS III Library of Fast Radiative Shock Models.}
   {The emission line spectra of the three regions are almost identical to each other and contribute to most of the emission line flux in the nuclear region. The emitting media contain multiple phases, the most luminous of which have temperatures ranging from $10^{4.8}$\ K to $10^{6}$\ K. Central photoionization models can reproduce some features of the spectra, but the fast radiative shock model provides the best fit to the data.}
   {The similarity between the three regions indicates that they belong to the same class of objects. Based on our comparisons, we conclude that they are shock regions located where the  jet of the active galactic nucleus impacts massive molecular clouds.}

   \keywords{Galaxies: individuals: NGC 1068 - Galaxies: active  - Galaxies: nuclei - ISM: clouds -  Infrared: ISM }

   \maketitle
   \footnotetext{The spectra of the three clouds B, C and D and the tables \protect\ref{table}, \protect\ref{table_comp_B}, \protect\ref{table_comp_C} and \protect\ref{table_comp_D} are available in electronic form at the CDS via anonymous ftp to cdsarc.cds.unistra.fr (XXX.XX.XXX.X) or via https:XXXX }

\section{Introduction}

The central region of NGC 1068 provides a unique opportunity to study some of the smallest spatial scales (1 arcsecond $\leftrightarrow$ 72 pc) in an active galactic nucleus (AGN) due to its proximity (14.4 Mpc) \citep{BlandHawthorn1997}. The galaxy's nearly face-on orientation offers a clear line of sight toward its central region, and its location near the celestial equator allows for observation from both hemispheres. As a result, NGC 1068 has been the subject of numerous publications since its original spectroscopic observation by Karl Seyfert in 1943 \citep{Seyfert1943}.


Observations of NGC 1068 at subarcsecond resolution from UV to radio wavelengths have been made possible by the use of space-based imaging, adaptive optics, and interferometry. These high-resolution observations have provided important insights into the physical processes occurring in the central region of this galaxy.

At gigahertz frequencies, a kiloparsec-scale ($\sim10$") radio jet oriented along a position angle of approximately 30 degrees is clearly visible in NGC 1068 \citep{Gallimore1996, Muxlow1996}. At smaller spatial scales \citep{Gallimore2004}, it appears that the jet is initially launched at a position angle of approximately 12 degrees from a compact radio source named S1, but it is bent approximately 0.3" north of it at the position of another compact radio source named component C (see Fig. \ref{images}). This deviation of the jet is proposed to be due to an interaction with an interstellar cloud \citep{Gallimore2004}.

At longer wavelengths, the molecular content of NGC 1068 is detected through CO emission lines, notably with ALMA \citep{GarciaBurillo2019}. The main feature of the central region is a massive circumnuclear molecular disk (CND) that is 400 pc wide and has a mass of $10^8\ \mathrm{M_{\odot}}$. The CND has a central hole of approximately 200 pc in diameter, within which the AGN is located, but not at the center. A smaller molecular disk with a mass of $3\times10^5\ \mathrm{M_{\odot}}$ has been observed at the position of S1, which has been identified as the mass reservoir and obscures the structure around the AGN \citep{GarciaBurillo2016, Imanishi2018, GarciaBurillo2019}. A molecular clump has also been observed north of S1, roughly at the bending point of the radio jet component C. High-resolution continuum observations in the ALMA spectral range have revealed a similar geometry to the observations made at gigahertz frequencies \citep{GarciaBurillo2019, Imanishi2020}.

In the IR, the details of NGC 1068 become accessible with adaptive optics-fed instruments. \citet{Gratadour2006} used deconvolution techniques on NAOS-CONICA observations to produce high angular resolution images of the central arcsecond in the K, L, and M bands. The nucleus S1 is the brightest source, particularly at short wavelengths. However, several compact sources with significant flux can be observed around it, particularly in the M band, where the northern sources are almost as bright as the nucleus. The positions of the two brightest sources, that is, the nucleus and IR1B, correspond to the positions of the radio components S1 and C, respectively.

At the shortest wavelengths, the central region of NGC 1068 has been imaged in the optical and UV wavelengths with the Hubble Space Telescope Faint Object Camera \citep{Macchetto1994}. The images of the source in the optical and UV continuum, as well as with the [O III] emission line filter, are very similar to each other on a large scale, revealing an extended $\sim$400 pc structure oriented along a position angle of approximately 30 degrees. The [O III] image revealed the complex filamentary structure of the narrow line region (NLR), which is an outflowing ionized bicone. Several bright emission line clouds were observed in the central arcsecond \citep{Crenshaw2000, Das2006}.

In this paper, we present a detailed study of the spatially resolved near-IR (NIR) spectra of three clouds (B, C, and D) from the inner NLR of NGC 1068 (named following the original convention from \cite{Evans1991}). In Section \ref{sec:obs}, we present the observations, followed by a quantitative description of the spectra in Section \ref{sec:results}. In Section \ref{sec:model}, we present the associated modelings, while a discussion of the nature of the objects is presented in Section \ref{sec:discu}. Finally, in Section \ref{sec:conlusions}, we summarize our conclusions.

\section{Observations and spectra extraction}
\subsection{Observations}
\label{sec:obs}
   \begin{figure}
   \centering
   \includegraphics[width=\linewidth]{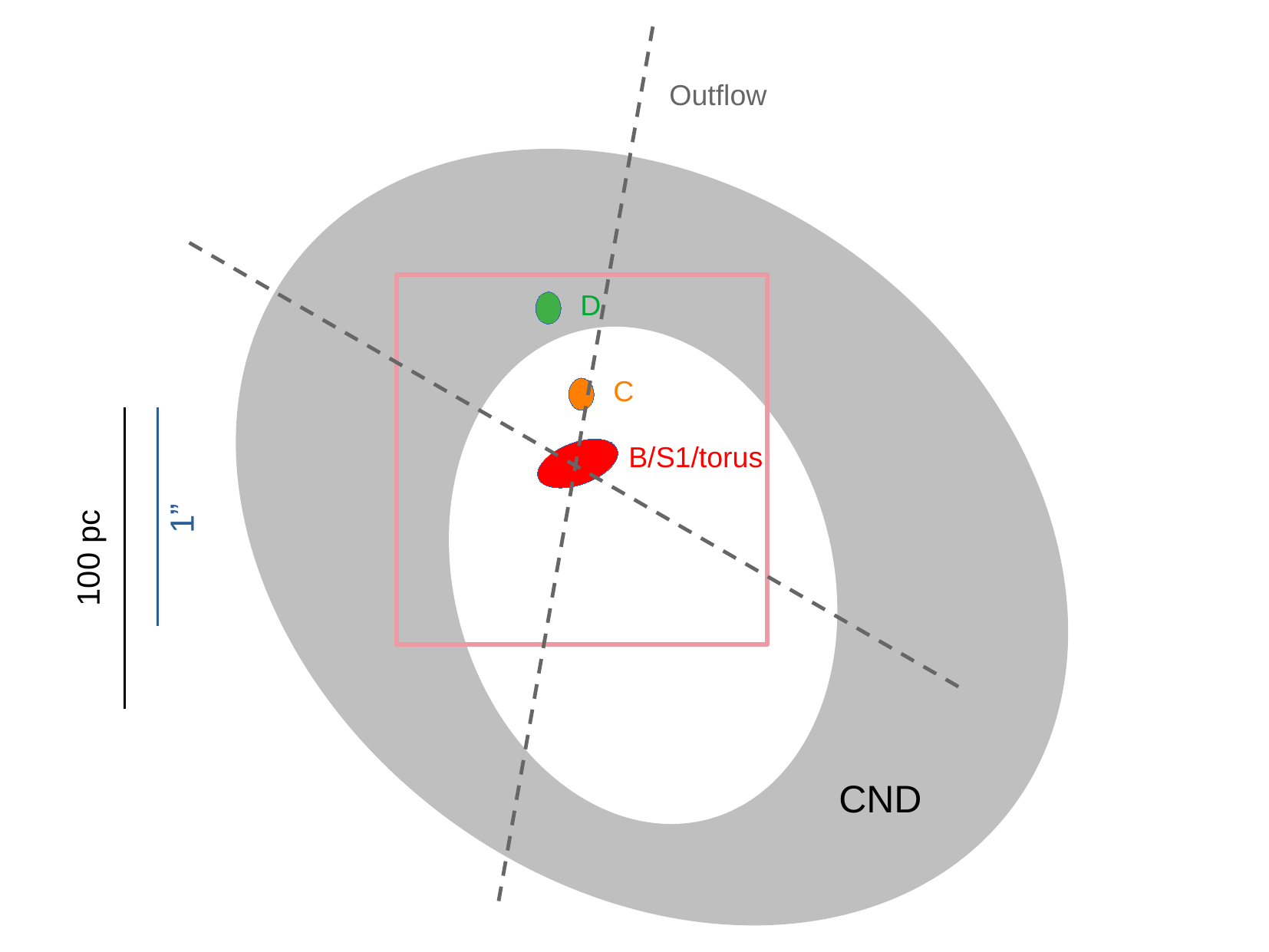}
   \includegraphics[width=\linewidth]{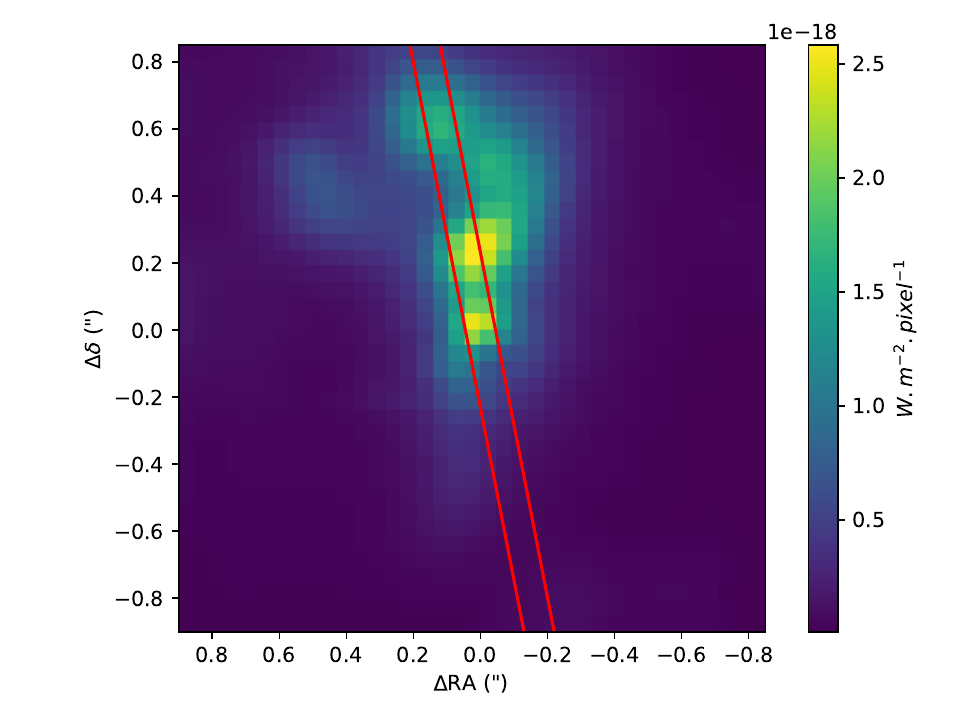}
      \caption{Top: Schematic representation of the main components in the central region of NGC 1068: the CND, the outer edges of the outflow bicone, and the three regions (B, C, and D) analyzed in this work. The field of view of the SINFONI observation is indicated with a pink square. Bottom: [Si VI] emission line map in the central region of NGC 1068 obtained with the SINFONI observation (pixel scale 50 mas) and position of the SPHERE slit (red lines, roughly aligned with the jet orientation). The coordinates are centered on the position of the K band continuum maximum. Following the naming convention from \cite{Evans1991}, the three regions studied in this paper are cloud B (in the center; at the same position as the radio component S1 and the compact torus), cloud C (0.25" north), and cloud D (0.7" north).
              }
         \label{images}
   \end{figure}
   
   \begin{figure*}
   \centering
   \includegraphics[width=\linewidth]{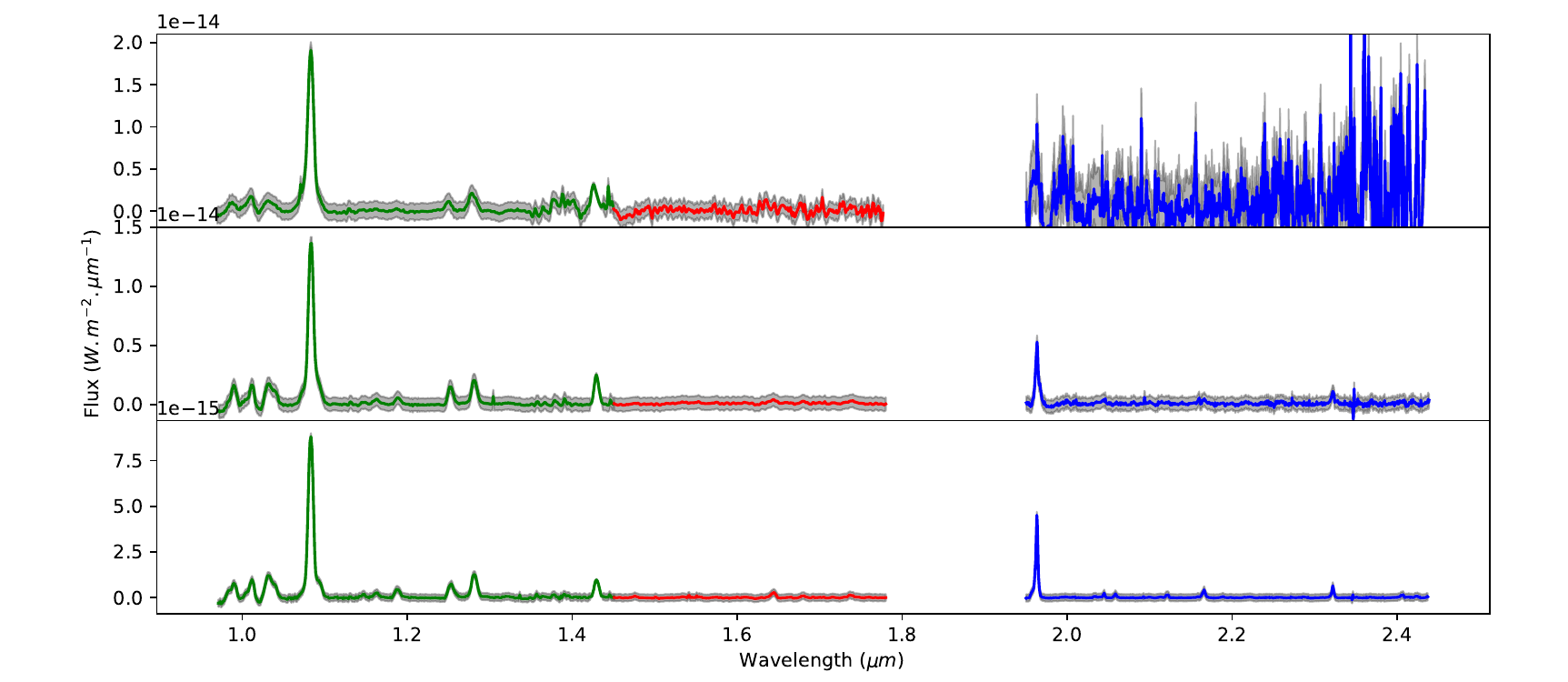}
      \caption{Combined YJHK spectra of the clouds with increasing distance from the nucleus. From top to bottom, the panels show a distance of  0.0" (B/S1/nucleus), 0.25" (C), and 0.7" (D). The SPHERE data are displayed in green, the SINFONI data in blue, and the averaged overlap in red. The estimated uncertainty is indicated as a gray area. }
         \label{spectra}
   \end{figure*}

This study is based on the merged analysis of two spectroscopic observations obtained with SPHERE/VLT and SINFONI/VLT. These observations are discussed individually in the following paragraphs.

\subsubsection{SPHERE}

We used a new observation performed with SPHERE in the IRDIS long-slit spectroscopy (LSS) mode \citep{Vigan2008, Beuzit2019} as part of the 0104.B-0242(A) ESO observing program (PI: J.-L. Beuzit). The observation was conducted under changing atmospheric conditions. The seeing was around 0.7" at the beginning of the observation (with a coherence time of the atmospheric turbulence between 5 and 6 ms) but quickly degraded to more than 1" (with a coherence time below 4 ms). As a result, the adaptive optics' performance varied a lot during the observation, providing a good resolution for the first few frames and quickly worsening. As a compromise between SNR and angular resolution, we selected the three best exposures of the object in terms of angular resolution and discarded the others.

The data reduction followed the instructions from the SPHERE manual\footnote{https://www.eso.org/sci/facilities/paranal/instruments/sphere/doc.html} and was performed as described in detail in \citet{Vermot2019}. We applied detector-level corrections, including dark subtraction, flat-field division, distortion correction, linear interpolation of the bad pixels, and realignment of the various exposures. As a second step, we performed spectroscopic calibrations, correction for the atmospheric transmission, and flux calibration from the 2MASS J and H magnitudes of BD-00413, the standard star observed during the run.

The final product is a flux-calibrated spectrum with a 100 mas angular resolution and an 8 nm spectral resolution. The slit covers 11" x 0.09", is centered on the maximum of continuum emission, and is oriented with $PA = 11\degree$ (see Fig. \ref{images}).

\subsubsection{SINFONI}

We used archival SINFONI integral field spectroscopy obtained as part of the observing program 076.B-0098(A). The instrumental setup provides spectrally dispersed images in the H and K bands with $R \sim 1500$ and $r = 100$ mas.

The data reduction was performed with the ESO pipeline in its standard mode, except for the flux calibration, which was performed a posteriori by matching the integrated HK band flux from the entire field of view to the one measured on 2MASS images of the galaxy. The image presented in Fig. \ref{images} corresponds to the [Si VI] continuum-subtracted emission line map obtained from this observation.

\subsection{Data interpolation and spectrum extraction}

To combine the two observations, we computed a synthetic long slit on the SINFONI data cube.  We used linear interpolation to rotate the cube in the spatial dimension around the maximum of continuum emission and extracted a two pixel-wide slit matching the orientation of the SPHERE one. Since this synthetic slit is 100 mas wide while the physical one from SPHERE is 90 mas, we applied a 0.9 correction factor to the SINFONI flux.

We extracted three spectra from each slit at positions corresponding to the maxima of emission lines: 0.0" (cloud B), 0.25" (cloud C), and 0.7" (cloud D). Each spectrum was summed over a 100 mas bin. The Y and J bands are only covered by SPHERE, the K band only by SINFONI, and the H band by both observations. For the latter, we interpolated the SINFONI data to the spectral resolution of SPHERE and took the average between the two observations. The match was perfect between the two datasets for clouds C and D, but the SINFONI spectrum of cloud B required recalibration in order to increase its flux by a factor two, which we attribute to a combination of imprecision in the synthetic slit extraction and differences between the PSFs of the two adaptive optics systems. Lastly, the continuum of emission was subtracted from the spectrum of each cloud to extract a pure emission line spectrum.

\section{Detected emission lines}

\label{sec:results}
The continuum-subtracted spectra of the three clouds are presented in Fig. \ref{spectra}, color coded by spectral domain. The green and blue parts correspond, respectively, to the SPHERE and SINFONI exclusive spectral zones, while the red part corresponds to the overlap of the two instruments. The spectra are presented, from top to bottom, with increasing distance from the nucleus. The uncertainty on the spectrum, indicated as a gray shaded area, was estimated by measuring the standard deviation of the spectra in regions free of emission lines. As can be observed, the SNR also increases with the distance from the center despite the flux of the emission lines decreasing. This is because the photon noise from the continuum of emission is at its maximum at the center. 

A significant result can already be deduced from the qualitative analysis of these spectra: They are remarkably similar to each other. They are dominated by a very strong emission line at $\sim 1.10\ \mu m$, to the left of which are three medium emission lines. Redward, two medium emission lines are found at $\sim 1.25\ \mu m$ and one at $\sim 1.45\ \mu m$, and  lastly, another strong emission line is found at $\sim 1.95\ \mu m$. This is valid for the spectra of clouds C and D, which are almost indistinguishable apart from their fluxes. Part of the spectrum of cloud B is unusable, but the emission lines in the J band are also fully consistent with the description above and with clouds C and D. This indicates that the three emission line regions are excited by the same mechanism, and most probably, the properties of the interstellar medium (ISM), for example, ionizing radiation field, density, and composition, are very close in these three clouds.

Altogether, we detected a few dozen emission lines, spanning a variety of elements and ionization energies. We present in Table \ref{table} the flux of all the emission lines detected with SNR > 3. We measured them with Gaussian fits (simple in the case of isolated lines and double in the case of overlapping lines). The uncertainties provided correspond to $1\sigma$, as estimated from the covariance matrix of the parameters. As discussed later in this paper, the identification of the lines has been done a posteriori by analyzing the output from the best CLOUDY models. We detected permitted emission lines, notably H I and He I. The He I line is the strongest emission line of the spectrum, and we detected up to seven emission lines from different hydrogen series. Assuming the Case B recombination case (clouds opaque to Lyman UV photons), we measured no extinction ($A_K \leq 0.15 $ for all three clouds at $3\sigma$). We also detected forbidden low-ionization energy emission lines, indicating the presence of a low-density environment. They are numerous and have relatively low fluxes. Lastly, we detected forbidden high-ionization energy emission lines, also known as "coronal emission lines." Two of them are of particular importance: [Si X] at 1430 nm, which has the strongest ionization energy among our lines (401 eV), and [Si VI] at 1960 nm, which is the second strongest line in our spectral domain.

\begin{table*}[ht]
\centering
\caption{\label{table}Summary of emission lines.}
\begin{tabular}{c|c|c|c|c|c}
Line & Rest & Ionization & Measured Flux & Measured Flux & Measured Flux \\
Identification & Wavelength & Energy & Cloud B & Cloud C & Cloud D \\
  & (nm)  & (eV) & $(10^{-18}\ W.m^{-2}$) & $(10^{-18}\ W.m^{-2}$) & $(10^{-18}\ W.m^{-2}$)\\
\hline
{[C I]} & 985.1 & 11.3 & 4.5 $\pm$ 1.4 & 0.7 $\pm$ 0.5 & 2.0 $\pm$ 0.3 \\
{[S VII]} & 986.9 & 281.0 & 7.8 $\pm$ 1.8 & 11.1 $\pm$ 0.8 & 5.1 $\pm$ 0.4 \\
{H I} & 1004.9 & 13.6 & 8.5 $\pm$ 2.0 & 3.0 $\pm$ 0.8 & 1.9 $\pm$ 0.5 \\
{He II} & 1012.4 & 54.4 & 11.4 $\pm$ 2.2 & 9.5 $\pm$ 1.2 & 5.5 $\pm$ 0.8 \\
{[S II]} & 1032.0 & 23.3 & 10.3 $\pm$ 0.9 & 13.1 $\pm$ 0.6 & 9.3 $\pm$ 0.4 \\
{[N I]} & 1039.8 & 14.5 & 5.3 $\pm$ 0.9 & 7.8 $\pm$ 0.6 & 4.6 $\pm$ 0.4 \\
{He I} & 1083.3 & 24.6 & 179.9 $\pm$ 7.6 & 109.2 $\pm$ 2.6 & 66.3 $\pm$ 0.9 \\
{H I} & 1093.8 & 13.6 & 7.1 $\pm$ 4.3 & 13.2 $\pm$ 1.5 & 6.1 $\pm$ 0.5 \\
{[P II]} & 1146.0 & 19.8 & 2.1 $\pm$ 0.5 & 1.5 $\pm$ 0.2 & 1.3 $\pm$ 0.1 \\
{He II} & 1167.3 & 54.4 & 3.2 $\pm$ 0.5 & 4.0 $\pm$ 0.3 & 2.1 $\pm$ 0.2 \\
{[P II]} & 1188.6 & 19.8 & 2.9 $\pm$ 0.5 & 4.9 $\pm$ 0.3 & 3.6 $\pm$ 0.2 \\
{[S IX]} & 1251.7 & 379.8 & 12.3 $\pm$ 0.6 & 11.3 $\pm$ 0.5 & 5.8 $\pm$ 0.2 \\
{H I} & 1281.8 & 13.6 & 24.8 $\pm$ 0.9 & 16.3 $\pm$ 0.6 & 10.4 $\pm$ 0.2 \\
{[Si X]} & 1430.2 & 401.4 & 24.1 $\pm$ 5.0 & 16.9 $\pm$ 0.5 & 6.6 $\pm$ 0.2 \\
{[Fe II]} & 1643.6 & 16.2 &  & 2.5 $\pm$ 0.2 & 2.0 $\pm$ 0.1 \\
{H I} & 1680.7 & 13.6 &  & 1.2 $\pm$ 0.2 & 0.7 $\pm$ 0.1 \\
{H I} & 1736.2 & 13.6 &  & 1.8 $\pm$ 0.2 & 1.1 $\pm$ 0.1 \\
{[Si VI]} & 1960.2 & 205.3 &  & 24.8 $\pm$ 1.7 & 15.1 $\pm$ 0.4 \\
{He II} & 2033.0 & 54.4 &  & 1.1 $\pm$ 0.2 & 0.5 $\pm$ 0.1 \\
{[Al IX]} & 2044.4 & 330.2 &  & 2.1 $\pm$ 0.2 & 0.9 $\pm$ 0.1 \\
{He I} & 2058.1 & 24.6 &  & 0.4 $\pm$ 0.2 & 0.9 $\pm$ 0.1 \\
{$H_2$} & 2121.2 & 0.0 &  & 0.7 $\pm$ 0.2 & 0.7 $\pm$ 0.0 \\
{H I} & 2165.5 & 13.6 &  & 1.8 $\pm$ 0.3 & 1.9 $\pm$ 0.1 \\
{[Ca VIII]} & 2322.2 & 147.3 &  & 4.3 $\pm$ 0.3 & 1.8 $\pm$ 0.1 \\
{H I} & 2402.9 & 13.6 &  & 1.8 $\pm$ 0.2 & 0.8 $\pm$ 0.0 \\
\end{tabular}
\end{table*}

\section{Physical conditions and ionization mechanism}
\label{sec:model}
\subsection{Goal and method}
In this section, we investigate the physical conditions and ionization mechanism of the clouds. To achieve this, we compare the observations with three ionization models: (1) a simple CLOUDY c17.02 \citep{Ferland2017} setup describing a cloud with constant density and temperature (without any ionization source; models PC1, PC2, and PC3); (2) a CLOUDY model of a cloud photoionized by a central AGN (model AGN); and (3) the MAPPINGS III shock model (which itself relies on CLOUDY; model SHK).

Instead of using a specific emission line ratio as a diagnostic between the different situations, we used the models to compute synthetic spectra with the same resolution as the observation and compared them directly to the observed spectra. This approach has several advantages over the use of emission line ratios. First, it takes into account every emission line. Second, it takes into account all the non-detections, which bring important constraints on the model. Third, it takes into account the absolute flux of the emission lines. Fourth, it makes no a priori assumptions about the identification of the lines. Moreover, overlapping close lines are naturally taken into account. Finally, by studying the best model, we can make an identification of the lines a posteriori.

We removed the [Fe II] lines from the models because they should be strong and ubiquitous according to our ionization models, but they are barely detected. The explanation for this weak [Fe II] signal is discussed in \ref{feII_disc}. Otherwise, synthetic spectra were computed and directly compared to the observations. For each model, we determined the best set of parameters by comparing the reduced chi-square between the model and the observation. 

\subsection{Physical conditions: Constant density and temperature}

To predict the emission line spectrum of a cloud with a uniform density and temperature on a large and fine grid, we first determined the physical conditions of the emitting regions by using CLOUDY c17.02. We call this model PC1. The grid covers a wide range of temperatures ($log(T/K)$ ranging from 3.70 to 9.00 with 0.02 steps) and densities ($log(n/cm^{-3})$ ranging from -5 to 9 with 0.05 steps).
We compared the spectra in absolute units of flux, assuming an emitting volume smaller than 125 pc$^3$.

The results for all three clouds are quite similar, with a deep minimum observed in the residual maps for temperatures ranging from $10^{4.2}$\ K to $10^{5.2}$\ K and densities of $100\ cm^{-3}$ or greater. The corresponding 2D maps for these parameters can be found in Fig. \ref{k2_density_temp}, while the 1D plots are presented in Figs. \ref{k2_density} and \ref{k2_temp}. The lower limit on density ($n \geq 10^{0.5}\ cm^{-3}$) is imposed by the maximum volume of the emitting cloud, but otherwise this parameter has little effect on the spectrum.

The best models shown in the top plots of Figs. \ref{cloud0_comps}, \ref{cloud1_comps}, and \ref{cloud2_comps} successfully reproduce the permitted emission lines (He I and hydrogen series) but do not produce any forbidden emission lines. The best-fit parameters for these models, including the corresponding emitting volumes, are provided in Table \ref{table_PC1}. The optimal temperature values for all three clouds are quite similar, ranging from 10$^{4.8}$ to 10$^{4.9}$\ K.

As indicated by Figs. \ref{k2_density_temp} and \ref{k2_temp}, a less likely secondary solution is also present in all three clouds, characterized by temperatures ranging from 10$^{5.5}$ to 10$^{6.5}$\ K. These solutions predict strong forbidden emission lines but weak permitted emission lines.

In the next step, we attempted to find a two-temperature combination that can reproduce the observed spectrum. To this end, we fixed the density at $n = 100\ $cm$^{-3}$ and computed the spectra for all possible temperature combinations, with flux ratios between the two components ranging from 0.05 to 0.45 in increments of 0.05. The results from this approach, which we call model PC2, are detailed in Table \ref{table_PC2}, with residual maps in Fig. \ref{k2_temp12}. For all three clouds, the majority of the emission is still attributable to a medium at $\sim 10^{4.8}$\ K (with acceptable temperatures ranging from 10$^{4.5}$ to 10$^{5.2}$\ K). Clouds B and C exhibit a secondary medium with temperature 10$^6$\ K (with corresponding flux ratios of 15\% and 25\%), while cloud D has a secondary medium with a temperature of 10$^{5.4}$\ K.

A comparison of the best PC2 models for clouds B, C, and D is presented in Figs. \ref{cloud0_comps}, \ref{cloud1_comps}, and \ref{cloud2_comps}. The results show that the best models for clouds B and C are successful in reproducing the permitted emission lines as well as most of the forbidden emission lines, but they fail to generate significant flux for the strong [Si VI] coronal emission line at 1960 nm. In contrast, the best model for cloud D accurately reproduces the He, H, and [Si VI] emission lines, but it fails to generate the correct flux for several other forbidden emission lines, specifically the coronal lines [S IX] and [Si X] at 1251 and 1430 nm. The best solution for clouds B and C is a good secondary solution for cloud D, and vice versa.

\begin{table}[ht]
\centering
\caption{\label{table_PC1}Best parameters for model PC1 (one temperature and one density).}
\begin{tabular}{c|c|c|c}
Cloud & Temperature & Density & Volume \\
 name & log(T/K) &  log(n/$cm^3$) & $pc^3$ \\
\hline
B & 4.9 & 2.35 & 0.06 \\
C & 4.84 & 1.65 & 1.28\\
D & 4.82 & 1.65 & 0.74 \\
\end{tabular}
\end{table}

\begin{table}[ht]
\centering
\caption{\label{table_PC2}Best parameters for model PC2 (two temperatures).}
\begin{tabular}{c|c|c|c}
Cloud & Temperature 1 & Temperature 2 & Flux ratio \\
 name & log(T/K) &  log(T/K) & - \\
\hline
B & 6.24 & 4.84 & 0.15 \\
C & 5.98 & 4.86 & 0.25\\
D & 5.38 & 4.74 & 0.30 \\
\end{tabular}
\end{table}

\begin{table}[ht]
\centering
\caption{\label{table_PC3}Best parameters for model PC3 (three temperatures).}
\begin{tabular}{c|c|c|c}
Cloud & $M_{4.8}$\ (Flux ratio) & $M_{5.4}$\ (Flux ratio) & $M_{6}$\ (Flux ratio) \\
 name & $M_{\odot}$ & $M_{\odot}$ & $M_{\odot}$ \\
\hline
B & 0.52 (0.63) & 0.44 (0.22) & 0.11 (0.15) \\
C & 0.37 (0.61) & 0.31 (0.21) & 0.09 (0.18) \\
D & 0.23 (0.62) & 0.23 (0.26) & 0.04 (0.12) \\
\end{tabular}
\end{table}

Each of the three clouds exhibits a multiphase ionized medium, with temperatures ranging from 10$^{4.8}$ to 10$^6\ K$. We created a third model (PC3) with three emitting media; the density fixed at $n = 100\ cm^{-3}$; and the temperatures set to $T_1 = 10^{4.8}\ K$, $T_2 = 10^{5.4}\ K$, and $T_3 = 10^{6}\ K$. We then determined the relative contribution of each medium to the observed spectrum. The best-fit parameters are presented in Table \ref{table_PC3}, and the corresponding spectra are shown in Figs. \ref{cloud0_comps}, \ref{cloud1_comps}, and \ref{cloud2_comps}. This model produces the best results, accurately predicting the flux of both permitted emission lines and the most important forbidden lines ([S IX], [Si X], and [Si VI]) for all three clouds. The ionized phase's mass is estimated to be approximately 1 $M_{\odot}$, 0.75 $M_{\odot}$, and 0.5 $M_{\odot}$ for clouds B, C, and D, respectively. In each case, the primary phase dominates at $T = 10^{4.8}$\ K (contributing to around 65\% of the flux), followed by $T\sim10^{5.4}$\ K (contributing to approximately 20\% of the flux) and $T\sim10^{6}$\ K (accounting for roughly 15\% of the flux).

\subsection{Central photoionization}

In a second model, we attempted to reproduce the observed spectra with a photoionization model of a cloud around a central radiation source. For this purpose, we used CLOUDY to simulate a cloud illuminated by an AGN. The luminosity of the source was relatively well constrained and set to $10^{38}$ W \citep{BlandHawthorn1997, Gallimore2001, Vermot2021}. The cloud's thickness was set to 10 pc, and its distance from the nucleus was 30 pc for clouds C and D, whereas for cloud B, it was set to 1 pc. In our simulations, we varied three parameters:
the density of the cloud, ranging from $10^{-2}$ to $10^{9}\ cm^{-3}$ in $10^{0.2}$ logarithmic steps; the temperature of the big blue bump (BBB), logarithmically ranging from $10^4$ to to $10^{6.5}$ K in logarithmic steps of $10^{0.1}$\ K (we note that the luminosity of ionizing photons is always $10^{38} W$); the optical to X-ray spectral index $\alpha_{OX}$, as defined in \citet{Zamorani1981} and implemented in CLOUDY with an explicit negative sign. Typical AGN have $\alpha_{OX} = -1.4$, while $\alpha_{OX}=0$ corresponds to the absence of X-ray emission.

With this model too, the residual maps are very similar from one cloud to another. An important degeneracy is observed between the temperature of the BBB and the $\alpha_{OX}$ parameter: the higher the temperature of the BBB, the lower the amount of  required X-rays (see Fig. \ref{AGN_k2_temp_dens}). In all cases, the model converges toward relatively low amounts of X-rays ($\alpha_{OX} \geq -1.2$). For all clouds, density is reliably constrained between 10$^{2.5}$ and 10$^{6.5}\ cm^{-3}$. For clouds B and C, the model converges toward a density $n \sim 10^4\ cm^{-3}$, while for cloud C, it converges toward $n \sim 10^{5.5}\ cm^{-3}$ (see Table \ref{table_AGN} for the exact values of the best parameters). 

Comparisons between the model and observation spectra for the three clouds are presented in Figs. \ref{cloud0_comps}, \ref{cloud1_comps}, and \ref{cloud2_comps}. For cloud D, the model correctly predicts the flux of the permitted emission lines as well as the strong [Si VI] coronal line, but it fails to reproduce the other weaker forbidden lines. For clouds B and C, the model predicts the presence of most of them, but it overestimates the flux of the [Ca VIII] emission line in the K band and the Hydrogen series, and it underestimates the [Si VI] line at 1960\ nm.

\begin{table}[ht]
\centering 
\caption{\label{table_AGN}Best parameters for model AGN (cloud photoionized by an AGN central source).}
\begin{tabular}{c|c|c|c|c}
Cloud & Temperature & $\alpha_{OX}$ & Density & Mass \\
 name & log(T/K) & - & log(n/$cm^3$) & $M_{\odot}$ \\
\hline
B & 5.0 & -0.8 & 3.6 &  $4.2 \times 10^{-4}$ \\
C & 4.8 & -0.9 & 4. & $1.1 \times 10^{-2}$ \\
D & 5.1 & -0.5 & 5.4 & $1.9 \times 10^{-2}$ \\
\end{tabular}
\end{table}

\subsection{Fast radiative shock}

\begin{table}[ht]
\centering 
\caption{\label{table_SHK}Best parameters for model SHK (fast radiative shock).}
\begin{tabular}{c|c|c|c|c}
Cloud & Velocity & Precursor & $\mathbf{B}$ & Shock surface \\
 name & $km.s^{-1}$ & $log(n/cm^{-3})$ & $\mu G$ & $pc^2$ \\
\hline
B & 1000 & 1 & 2.7 & 0.32 \\
C & 1000 & 1 & 5.0 & 0.25\\
D & 975 & 2 & 20. & 0.015\\
\end{tabular}
\end{table}

In the final modeling attempt, the observed spectrum was modeled as emerging from a region ionized by a shock using the MAPPINGS III Library of Fast Radiative Shock Models \citep{Allen2008}. The model assumes that the shock front generates a strong radiation field of extreme ultraviolet and soft X-ray photons, leading to significant photoionization in the shock itself and in the region ahead of it. The model varies three parameters: the velocity of the shock, ranging from 100 to 1000 km.s$^{-1}$ in steps of 25 km.s$^{-1}$; the preshock density, ranging from 10$^{-2}$ to 10$^3$ cm$^{-3}$ in logarithmic steps of 10$^1$;
the preshock transverse magnetic field $\mathbf{B}$, with values ranging from 10$^{-4}$ to 10$^3$ $\mu$G, in an irregular sampling covering the extremes expected in the ISM while also sampling more finely the magnetic field values near equipartition.

The library does not explore a full regular parameter space, as the sampling of the magnetic field (number of samples and selected values of $B$) differs from one value of the density to another. Consequently, 2D maps of the chi-square as a function of the parameters are sparse and difficult to read, and thus we only present1D projections (minimum value of the chi-square for a given parameter). These are available in Figs. \ref{shock_k2_velocity}, \ref{shock_k2_density}, and \ref{shock_k2_mag}. They favor fast shocks ($v \geq 750\ km.s^{-1}$) and exclude the lowest densities ($n \geq 10^{-1}\ cm^{-3}$) and magnetic fields ($\mathbf{B} \geq 10^{-1}\ \mu G$).

As with the previous models, the optimal parameters of the three clouds are very similar to each other. The minimum value of the chi-square is obtained for the highest velocity ($v_s \sim 1000$\ km.s$^{-1}$), low densities ($n \sim 1 - 100$\ cm$^{-3}$), and intermediate magnetic fields ($B \sim 1 - 100\ \mu$G). The exact best parameters for each cloud are presented in Table \ref{table_SHK}.


For the three clouds, the best model is in excellent agreement with the observation. All the main emission lines are correctly predicted by the model (permitted and forbidden) with fluxes very close to what is observed (comparisons between the spectra are presented in the bottom plots of Figs \ref{cloud0_comps}, \ref{cloud1_comps}, and \ref{cloud2_comps}). The ratio between the strong He I and [Si VI]$_{1960}$ emission lines is well reproduced, as well as the flux of the other coronal lines, [S IX] and [Si X]. The flux of the hydrogen emission lines is overpredicted, but the shape of the complex emission features redward of 1100\ nm is remarkably well reproduced. 

The best model corresponds to a shock at 1000\ km.s$^{-1}$, which is the highest velocity of the MAPPINGS model. As a consequence, it is possible that higher shock velocities could provide an even better fit to the data.

\begin{figure*}
\centering
\includegraphics[width=\linewidth]{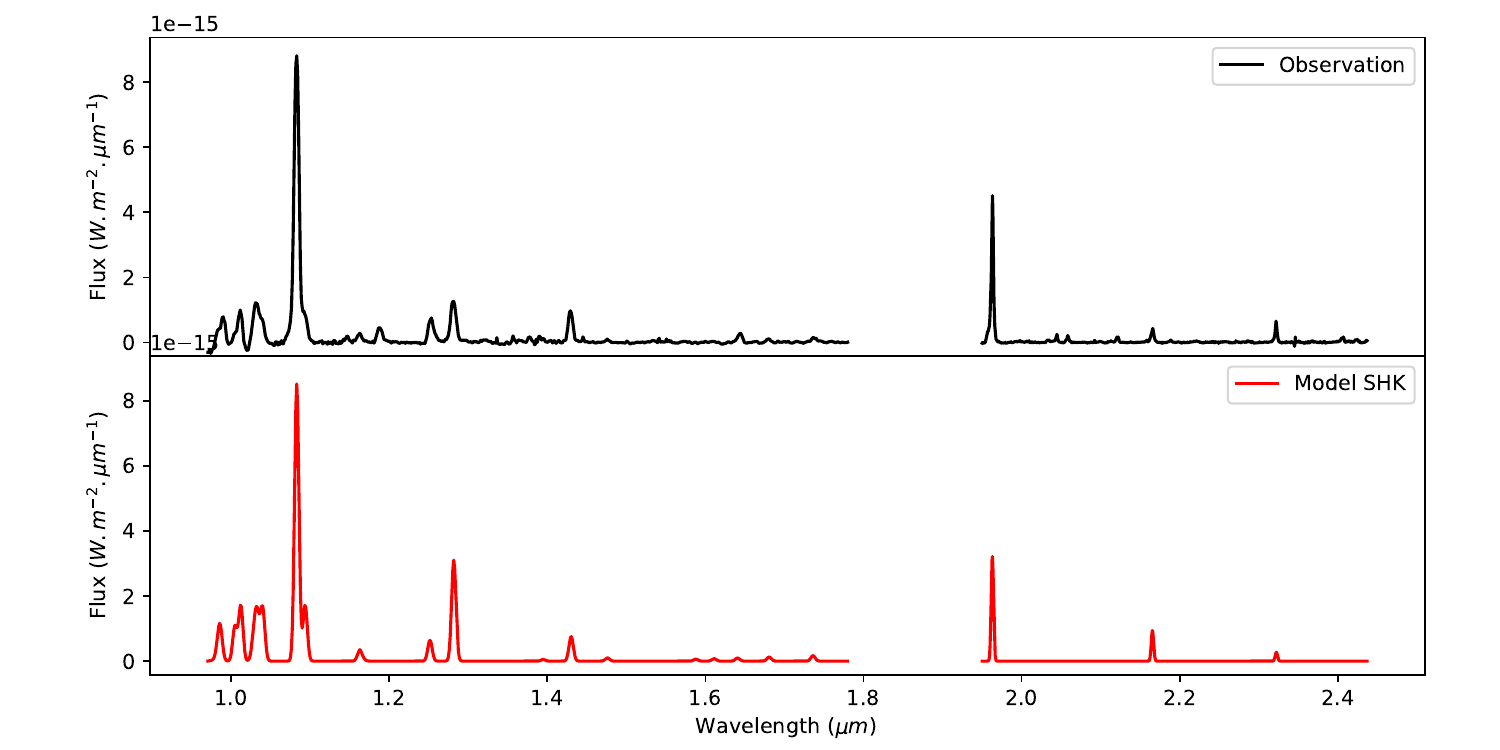}
  \caption{Cloud D: Comparison between the observed spectrum (top, black) and the best model (bottom, red). The numerical values of the fluxes from the model's main emission lines are compared with those from the observation in Table \ref{table_comp_D} from the appendix.}
  \label{cloud2_shock_comp_specs}
\end{figure*}

\section{Discussion}
\label{sec:discu}
\subsection{Similarities between the clouds}
One of the most striking results of the analysis presented in this paper is the strong similarity between the emission line spectra of the three objects, which leads to an almost identical modeling. Not only are the ratios of the various emission lines extremely similar between them, but their absolute fluxes are also of the same magnitude. We conclude that they belong to the same class of object, with similar physical conditions created by the same ionization mechanism.

\subsection{Ionization mechanism}

\subsubsection{Comparison between the models}

\begin{table}[ht]
\centering
\caption{\label{table_chi2}Reduced chi-square for the various models.}
\begin{tabular}{c|c c c|c c}
Cloud & PC1 & PC2 & PC3 & AGN & SHK \\
\hline
B & 11.9 & 8.8 & 8.8 & 10.5 & 10.6 \\
C & 9.8 & 4.4 & 1.9 & 5.6 & 4.1 \\
D & 22.1 & 10.0 & 2.8 & 9.4 & 7.9 \\
\end{tabular}
\end{table}

It is clear from the comparison to the models that the emitting medium is multiphase, with temperatures ranging from $T_1 = 10^{4.8}\ \mathrm{K}$ to $T_3 = 10^{6}\ \mathrm{K}$. The detection of strong forbidden lines with ionization energies as high as 400 eV imposes the highest temperatures, but a lower energy phase must be present to explain the strong emission from non-fully ionized elements.

Photoionization models of a central source with a simple blackbody spectral energy distribution fail to reproduce the observed spectra, as they can only reproduce one of the above-mentioned phases, depending on the temperature of the central source. More complex photoionization models with a source combining UV and X-ray contributions can produce a multiphase medium when illuminating dense clouds, reproducing the observed He I flux as well as some forbidden lines. Nevertheless, this AGN central photoionization model fails to reproduce some features of the observed spectrum, such as the complex structure blueward of He I or the ratios between the many high-energy forbidden lines.

The last model presented in this paper, which assumes that the emission lines are stimulated by a fast radiative shock, provides a remarkably good visual fit to the data and can reproduce all the essential features of the spectra (see Fig. \ref{cloud2_shock_comp_specs} and Tables \ref{table_comp_B}, \ref{table_comp_C}, \ref{table_comp_D}). Despite significant differences between the measured emission fluxes and those predicted by our model, the reduced chi-square value we presented remains relatively low, which can appear surprising. This is because it is calculated across all pixels in the spectrum, incorporating the many non-detected emission lines in the process, which effectively lowers the value of the reduced chi-square. Furthermore, we employed a meticulous Gaussian fitting method to measure the fluxes reported in Tables \ref{table_comp_B}, \ref{table_comp_C}, and \ref{table_comp_D} that yields more precise flux values than a simple sum of the corresponding pixels. 

Given the presence of an AGN in the vicinity of the emitting media, it is reasonable to assume that fast shocks could be responsible for the excitation of the clouds. Therefore, we conclude that this model is the most plausible one to explain the observed emission lines. For clouds C and D, the chi-square analysis significantly favors this shock model over the central photoionization one (see Table \ref{table_chi2}). For cloud B, the two models have a similar chi-square since the K band spectra is too noisy to discirminate between the two.

Our best fit analysis reveals that the shocks must be at high velocity (close to or exceeding 1000 $\mathrm{km.s^{-1}}$); the preshock density is relatively high, in the range of $1-10$ cm$^{-3}$; and the magnetic field is near equipartition. These parameters provide insights into the structure of the medium, which is composed of three regions: the shock itself, which heats the gas up to temperatures of $10^7\ \mathrm{K}$ and generates strong ionizing radiation; a preshock region that is ionized by this radiation and forms an HII region; and a cooling recombination region located behind the shock. For a quantitative description of the ionization, temperature, and density profiles around a shock with similar parameters ($v_s = 1000\ \mathrm{km.s^{-1}}$, $n = 1\ \mathrm{cm^{-3}}$, $\mathbf{B} = 3.23\ \mathrm{\mu G}$), we refer to Figs. 4 and 8 in \citet{Allen2008}.

\subsubsection{Improved diagnostic diagram}

"Baldwin, Phillips and Terlevich" (BPT) diagrams \citep{BPT1981, Veilleux1987, Kewley2001} are a powerful tool using emission line ratios to distinguish between an AGN and star formation activity from optical spectra. In a series of recent publications, \citet{Dagostino2019a, Dagostino2019b} proposed an improved 3D version of this type of diagram that takes into account the measured velocity dispersion derived from the lines as well in order to distinguish between shocks, AGN activity, and star formation and applied it to NGC 1068. The observation, obtained with the Wide Field Spectrograph \citep[WiFeS;][]{Dopita2007, Dopita2010}, is at a lower spatial resolution than ours and does not allow the resolution of individual clouds in the central arcsecond to be obtained. However, their results appear consistent with ours, highlighting that the entire central region (400 pc <=> 6") is dominated by shocks \citep[Fig. 9 in][]{Dagostino2019b}.

In our work, the conclusion is purely based on the analysis of the fluxes of various emission lines, while in their work, the decisive argument is the measured velocity dispersion of the emission lines; when high, this is a strong indicator of shock excitation \citep{Rich2011, Ho2014}. We cannot directly apply their method to our observation since we are observing other emission lines and do not have enough SPHERE spaxels to compute a proper emission line ratio function. However, despite our lower spectral resolution, we can attempt to measure the velocity dispersion from the emission lines in the SINFONI dataset for clouds C and D.

The resolution of our observation is R $\sim$ 1500, corresponding to an expected full width half maximum (FWHM) of 1.35 nm at 2 um for an unresolved emission line. For our strongest emission line, [Si VI] at 1965 nm, we measured an FWHM of 4 nm for cloud C and 3 nm for cloud D, corresponding to intrinsic velocity dispersions of 245 km s$^{-1}$ and 150 km s$^{-1}$, respectively. For [Ca VIII], we found similar values: 220 km s$^{-1}$ and 130 km s$^{-1}$. These values are not as high as the highest ones reported in \citep{Dagostino2019a, Dagostino2019b} for the central region, but they still correspond to the beginning of the shock-dominated sequence from the 3D diagram, especially if we consider that a) a higher velocity dispersion component could be blended in our line and b) spatial mixing could be present in their lower spatial resolution IFU observation. Overall, we consider that our results are fully consistent with theirs, strengthening the interpretation that the excitation of the emission lines in the central region of NGC 1068 is dominated by shocks.

\subsubsection{[Fe II] emission lines}
\label{feII_disc}
We noted an unexpected lack of [Fe II] line emission in our spectra. According to the photoionization and shock models used to describe the other emission lines, these lines should be prominent. This discrepancy led us to remove [Fe II] lines from our analysis. In this section, we explore why [Fe II] might be so weak.

In the ISM, UV and X-ray observations often find less iron than expected. Most of it, about 90\%, is thought to be hidden in dust particles \citep{Dwek2016, DeCia2018}. One might initially guess that iron is similarly depleted in our case, thereby accounting for the weakness of [Fe II] emission. However, our best explanation for the emission lines we do see involves a fast-moving shockwave traveling at 1000 km s$^{-1}$. Such a shockwave would destroy dust, freeing the iron. Moreover, if iron were depleted, we should also miss signatures of other elements found in dust, such as silicon, but we observed strong silicon lines. A closer look revealed that the [Si X] and [Si VI] lines we observed come from ionization states requiring extremely high temperatures (at least $10^5$\ K and $10^6$\ K, respectively), whereas [Fe II] lines are typically found at temperatures of $\sim 10^4$ K or less, as shown in Figs. \ref{feII_temp_dens}, \ref{SiVI_temp_dens}, and \ref{SiX_temp_dens}.

In the MAPPINGS shock model, there are two places where temperatures are below $10^4$\ K: the gas that has not been hit by the shockwave yet and the cooling zone behind the shock \citep[see Fig. 5 in][]{Allen2008}. The model predicts that the vast majority of the [Fe II] emission should arise in the cooling zone, due to the higher $n_H$ density. So we need to explain why there is no [Fe II] there. We think there are two possibilities.

One possibility is that dust forms again after the shock. If dust particles form quickly in the cooling region, they could capture a lot of the iron, explaining why we do not see [Fe II]. The same would happen to silicon, but the [Si X] and [Si VI] lines would still be visible, as they come from the hot, shocked region. Dust formation might indeed be possible here, as there are enough atomic elements to form dust (e.g., Fe, Si, C, which are released from the destroyed dust); the shock would make the medium dense; and the dust wind from the central black hole could provide "seeds" for dust formation. We note that [C I], which is a low-ionization energy state coming from an element constitutive of dust (similar to [Fe II]; see Fig. \ref{CI_temp_dens}), is also overpredicted by the shock model (see Tables \ref{table_comp_B}, \ref{table_comp_C}, and  \ref{table_comp_D}), which supports this hypothesis. As a possible follow-up observation, the nondetection of the [Si II] line at 34.85\ $\mu m$ could confirm it.    

The second possibility is that no low-temperature area exists after the shock. Due to intense radiation from the AGN central engine, it is possible that the cooling region behind the shock never cools to below $10^4$\ K. Additionally, the central region of NGC 1068 is known to have recently experienced star formation \citep{Storchi2012, Vermot2019, Rouan2019}, so hot stars in the nuclear star cluster (or directly inside the clouds) could contribute to the heating of the ISM in the post-shock region, for instance,  in the Orion HII region, which exhibits low [Fe II] emission \citep{walmsey2000}. In that case, we would not see any [Fe II] because iron cannot emit it in those conditions.

To confidently choose between these two options, we would need more complex simulations. However, this is not within the scope of our current work.

\subsection{Nature of the objects}

\subsubsection{Other properties}

In addition to the emission line spectra, the objects have been detected in the mid-IR \citep{Gratadour2006}, forming a structure very similar to the one observed
in the [Si VI] image presented in Fig. \ref{images}. Cloud B corresponds to the nucleus, cloud C to IR-1B, and cloud D to the superposition of IR-3 and IR-4, as named in \citet{Gratadour2006}. The other clouds detected in the mid-IR also have counterparts in the [Si VI] image. While the secondary clouds are much less luminous than the nucleus in the K band, this difference in flux becomes negligible at longer wavelengths, such as the M band, where most of the energy is radiated: $M_{nucleus} = 6.6$, $M_{IR-1B} = 6.7$, $M_{IR-3+IR-4} = 7.6$. Therefore, all three objects can be considered strong sources of IR radiation.

The three clouds also have molecular counterparts. In \citet{MullerSanchez2009}, the authors identified several molecular structures in SINFONI observations through an $H_2$ rovibrational line. Clouds B and C correspond to the structures named southern and northern tongues, respectively, identified in the $H_2$ observations. The southern tongue is on the line of sight of the IR peak, and cloud D is located within a larger CO molecular structure, which is an overdensity of the CND. The dynamical modeling of the southern and northern tongues indicates that they could be molecular streamers fueling the nucleus. The mass of these streamers ranges from a few $10^6$\ to several $10^7$\ $\mathrm{M_{\odot}}$, estimated by various methods. The CO counterpart to component B is the molecular disk observed at the position of S1, with a mass estimated at $3 \times 10^5\ \mathrm{M_{\odot}}$. A small CO clump is observed north of it at the position of cloud C, and cloud D lies again within the larger CND. In our [Si VI] observation of the objects, a velocity gradient can be measured along clouds C and D (see Appendix \ref{app:velocities}; cloud B is too noisy). The amplitude of the velocity gradient is similar between the two clouds, and if interpreted as gravitational rotation, it would correspond to masses greater than a few $10^6\ \mathrm{M_{\odot}}$.

The three clouds may also correspond to the radio sources detected in \citet{Gallimore2004} between 1 and 10 GHz. However, aligning the radio image with the NIR emission is uncertain due to the high uncertainty in absolute astrometry \citep{Capetti1997} and the dissimilarities in morphology between the two images that prevent cross-registration. Despite this, if S1 and the IR emission peak are associated with the nucleus \citep{Gallimore2004, Gratadour2006, GamezRosas2022}, cloud B may correspond to the radio component S1 and cloud C to the radio component C. Cloud D could be associated with component NE, but with an offset of about 0.05". At these wavelengths, the similarities between the objects are striking, particularly between S1 and C, which have comparable fluxes, spectral slopes, and geometries \citep[see Figs. 3, 4, and 5 in][]{Gallimore2004}. This comparison also holds for component NE at 5 and 8.4 GHz, but not at 1.4 GHz, where it appears more extended and luminous.

Thus, the similarities observed in the NIR emission line spectra of the clouds extend to other wavelengths. All three clouds are strong sources of IR radiation, indicating the presence of hot dust. They also have strong molecular counterparts and are massive, likely with masses greater than or equal to $5 \times 10^6\ \mathrm{M_{\odot}}$. Finally, energetic phenomena occur in their vicinity, as evidenced by the strong radio continuum.

\subsubsection{Shocked molecular clouds}

Our new results indicate that clouds B, C, and D are giant molecular clouds shocked by the AGN jet, as evidenced by the high-energy forbidden lines observed in our NIR spectra and the nearby strong radio sources. The presence of hot dust and molecular content indicates that these clouds have not been completely destroyed by the jet yet, and they could be transient structures.

If the IR peak indeed traces the position of the nucleus \citep{GravityCollab2020, Vermot2021, GamezRosas2022}, then cloud B is in the line of sight of the nucleus. If it is interacting with the jet, as evidenced by this work, it means that it has reached the actual position of the nucleus. As such, cloud B could simultaneously be the mass reservoir for the current accretion episode and the dense obscurer giving the nucleus its type 2 properties. Both the hot phase of the molecular content observed with SINFONI through the $H_2$ rovibrational line \citep{MullerSanchez2009} and the colder phase traced by various ALMA molecular lines \citep{GarciaBurillo2019, Impellizzeri2019, Imanishi2020} point toward complex kinematics that are incompatible with a simple rotation profile and probably result from the superposition of rotation, counter rotation, and outflow kinematics. Thus, if cloud B is indeed currently fueling and obscuring the nucleus, it has not reached a steady orbital state around the central mass.

Cloud D is part of the CND, and clouds E and F (not discussed in this paper) are also part of the CND, providing evidence that the CND shocked region is extended. The presence of strong IR emission and molecular tracers at the position of the clouds reveal that the interaction with the jet is not sufficient to destroy the densest molecular clouds of the CND.

Cloud C is located in the empty region between the CND and the nucleus, and it appears to be an extended structure as shown in the high angular resolution [O III] HST/FOC image \citep{Macchetto1994} and the $H_2$ rovibrational line image \citep{MullerSanchez2009}. A kinematics analysis in he latter reference indicates that the structure is in a highly elliptical orbit around the nucleus and is streaming toward it. The associated radio component C is located on its eastern edge, which is the leading side of its orbit, as per \citet{MullerSanchez2009}. Our results indicating that the emission lines come from a shocked region are consistent with the scenario of cloud C being a tidally disrupted molecular cloud that is shocked and ionized as its orbit brings it into contact with the outflow.

In this scenario, clouds D, C, and B would chronologically trace the evolutionary stages of the gas reservoir for the AGN activity. The CND (cloud D) would constitute the bulk of the molecular reservoir. Its interaction with the outflow would result in a loss of angular momentum, causing some clouds to detach from it and fall into the gravitational potential (cloud C). Ultimately, these clouds would reach the nucleus (cloud B) and trigger the accretion onto the supermassive black hole (SMBH), potentially obscuring it in the process.

\begin{figure}
\centering
\includegraphics[width=\linewidth]{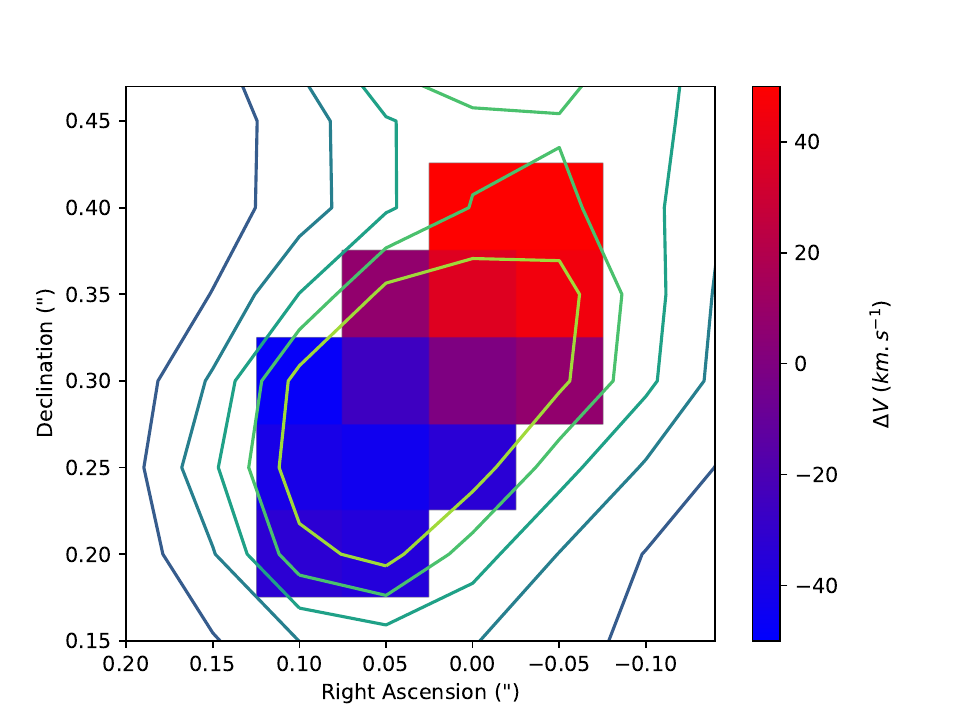}
  \caption{Cloud C: Velocity map based on the measured Doppler shift of the [Si VI] emission line. The contours represent flux levels.}
  \label{cloud1_velocity_txt}
\end{figure}

\subsubsection{Cloud C: A potential secondary AGN}

Cloud B is located at the position of the IR peak, which is reliably associated with S1 \citep{GravityCollab2020, GamezRosas2022}, the AGN nucleus \citep{Gallimore2004}. Given the similarities observed between clouds B and C in their NIR emission lines, radio flux and spectrum, and IR flux, it is reasonable to consider that cloud C could also host an active SMBH.

In the IR images presented in \citet{Gratadour2006}, the authors discuss the challenge of explaining the high temperatures observed in IR-1B with radiation from the central source and suggest that the interaction of the jet with its environment or the presence of very small grains could account for it. However, the presence of an active SMBH in cloud C would provide a simpler explanation for this temperature excess.

While component S1 is identified as the nucleus in \citet{Gallimore2004}, due to an inverted spectrum below 5 Ghz and a flat spectrum above as well as the presence of bright $H_2O$ masers in a rotating pattern, component C is not identified as a nucleus despite having an almost identical spectrum and flux as S1 and the presence of $H_2O$ masers (without a rotating pattern). Component C is instead associated with the interaction between the jet and a molecular cloud, due to the non-exact alignment with optical emission lines and the apparent deviation of the jet at this position. However, recent observations by \citet{Morishima2023} have revealed a ring-like structure of $H_2O$ masers around component C, which could be indicative of rotation. While this does not rule out the possibility that component C results from the interaction of the jet with a molecular cloud, the authors suggest that it could also indicate the presence of a rotating disk around a $\sim 10^6\ \mathrm{M_{\odot}}$ SMBH. This mass estimate is consistent with the velocity gradient measured on the [Si VI] emission line in our SINFONI data (see Fig. \ref{cloud1_velocity_txt}).

At radio wavelengths, the properties of components S1 and C in NGC 1068 are comparable to those of components N1 and S in NGC 6240 in terms of flux, geometry, and (to a lesser extent) spectral slope \citep{Gallimore2004_6240}. In the case of NGC 6240, these two radio sources are identified as a double AGN. The SINFONI observation of NGC 6240 \citep{Ilha2016} also shows the peak of the coronal emission lines at the position of the two AGN, closely but not perfectly lined up, similar to NGC 1068. The two nuclei of NGC 6240 are also strong sources of NIR to mid-IR radiation \citep{Max2005, Mori2014}. 

The decisive argument to classify NGC 6240 as a double AGN was the association of the two strong IR and radio sources to two hard, luminous X-ray nuclei discovered with Chandra \citep{Komossa2003}. However, in the case of NGC 1068, clouds B and C are too close together for X-ray telescopes to resolve them. Nevertheless, the Chandra image presented in \citet{Young2001} reveals an extension of the nuclear component in the direction of cloud C, and the spectral modeling done in the same paper highlights the difficulty in fitting the X-ray spectrum with a single hot plasma model. These elements are consistent with the presence of a secondary source.

In essence, we contend that even without S1, the central region of NGC 1068  would still be classified as an AGN based on the properties of cloud C, such as its strong inverted radio continuum, its ring-like distribution of masers, and its strong IR emission. Our results corroborate this interpretation, as they reveal strong similarities between the emission line spectra of clouds B and C, with the former being identified as an AGN. However, this argument is not conclusive since the shocked atomic content detected in our study could also arise from a shock with the jet. Given that cloud C is in the outflow region, this hypothesis is favored until further evidence for AGN activity in cloud C (such as resolved X-ray emission) is found. We plan to investigate this question further in the near future.

\section{Conclusions}

\label{sec:conlusions}
The combination of SPHERE and SINFONI data enabled us to produce a comprehensive YJHK spectrum for three emission line regions named B, C, and D in the central arcsecond of NGC 1068. Our analysis reveals that these regions exhibit similar emission line spectra, suggesting that they belong to the same class of objects. The various modeling techniques utilized in this study produced consistent results for all three objects.

Our investigation involving the use of CLOUDY to produce synthetic emission line spectra for several models and the comparison of these spectra with the full observations revealed the presence of multiple phases in the clouds. The temperatures of these phases range from $10^{4.8}$\ K to $10^6$\ K. Our findings indicate that a central AGN photoionization model is not able to accurately reproduce the variety of emission lines observed in the spectra. However, the MAPPINGS III fast radiative shock model is able to reproduce the observed emission line spectra well, with shock velocities of $v_s \sim 1000\ \mathrm{km.s^{-1}}$ and near equipartition magnetic fields.

Our study suggests that the three objects are likely to be giant molecular clouds that are either orbiting or streaming toward the nucleus and are being shocked by the same radio jet as they enter the outflowing bicone. It is possible that most of the properties associated with the nucleus and its torus, such as mid-IR and radio continuum emissions as well as atomic and molecular lines, originate from cloud B, which is coincident with the nucleus and may be currently feeding the central SMBH. However, given the similarities between clouds B and C, together with their shared AGN properties, such as high IR flux, radio continuum, $H_2O$ masers, and $M_{dyn} \gg 10^6 \mathrm{M_{\odot}}$, it is also possible that both objects host an AGN, with cloud C being a secondary AGN orbiting cloud B.

\begin{acknowledgements}
We thank Eric Lagadec and the SPHERE consortium for carrying out this observation as part of the GTO \textit{Other Science} program. This work was made possible by the support of the international collaboration in astronomy (ASU mobility) with the number $CZ.02.2.69/0.0/0.0/18\_053/0016972$ and the institutional project RVO:67985815. ASU mobility is co-financed by the European Union. 
\end{acknowledgements}

\bibliography{biblio}

\begin{appendix} 

\begin{figure*}[!b]
    \includegraphics[width=\linewidth]{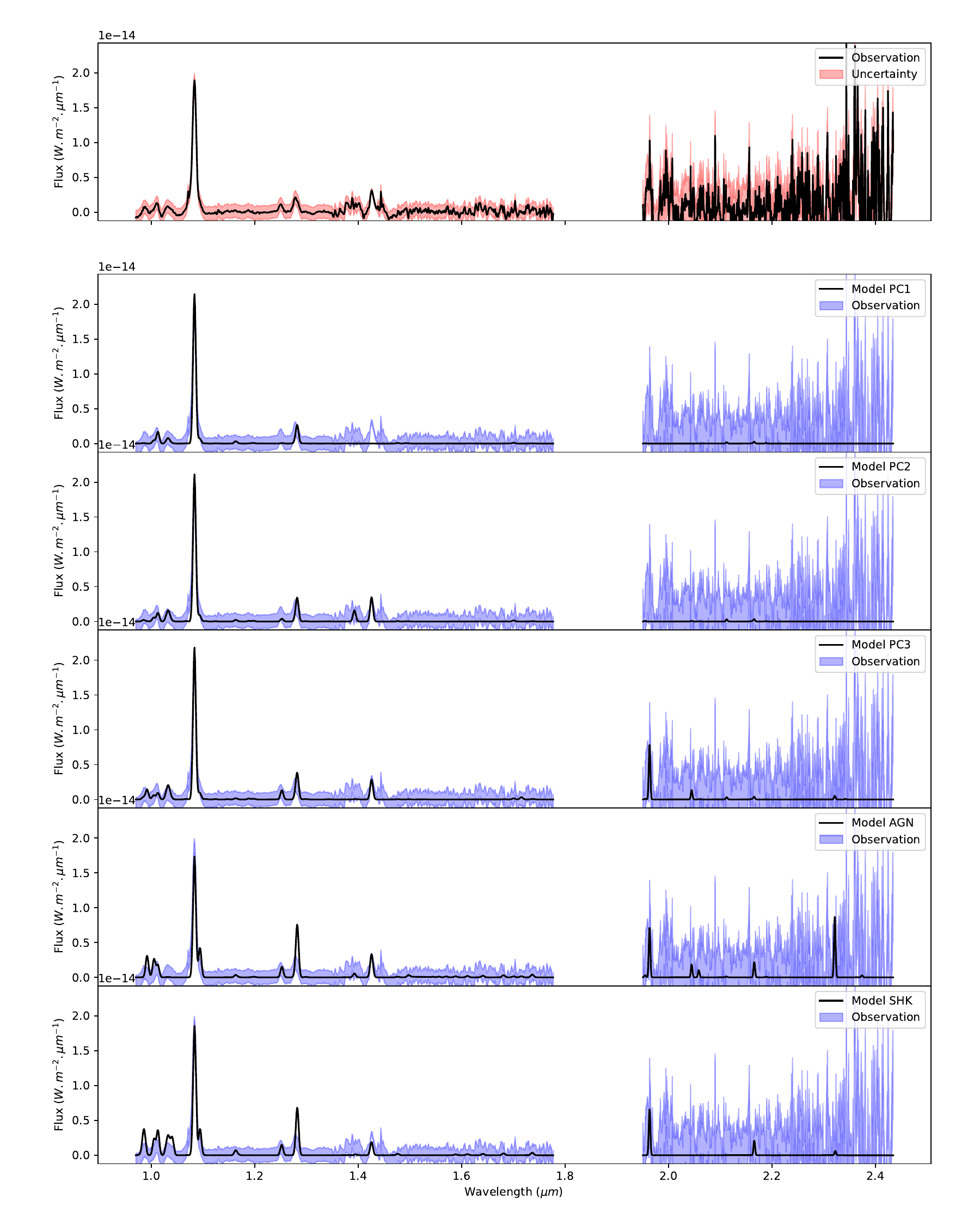}
  \caption{Cloud B: Comparison between the observed spectrum and the various models. The top panel shows the observed spectrum in black with the estimated uncertainty overlaid in transparent red. The five following plots show the prediction from models PC1, PC2, PC3, AGN, and SHK, in that order, with the observation overlaid in transparent blue.}
  \label{cloud0_comps}
\end{figure*}

\begin{figure*}[!b]
    \includegraphics[width=\linewidth]{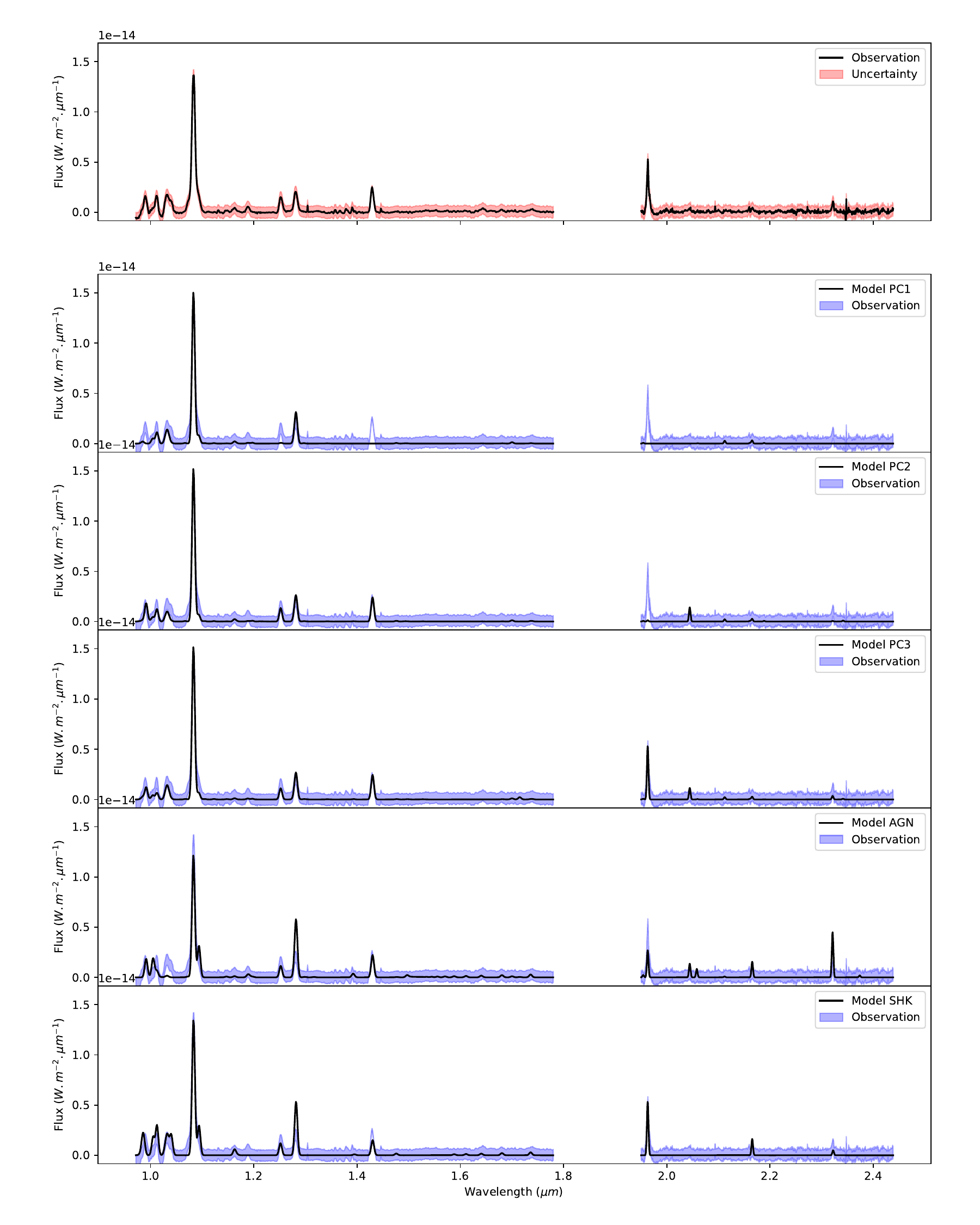}
  \caption{Cloud C: Comparison between the observed spectrum and the various models. The top panel shows the observed spectrum in black with the estimated uncertainty overlaid in transparent red. The five following plots show the prediction from models PC1, PC2, PC3, AGN, and SHK, in that order, with the observation overlaid in transparent blue.}
  \label{cloud1_comps}
\end{figure*}

\begin{figure*}[!b]
    \includegraphics[width=\linewidth]{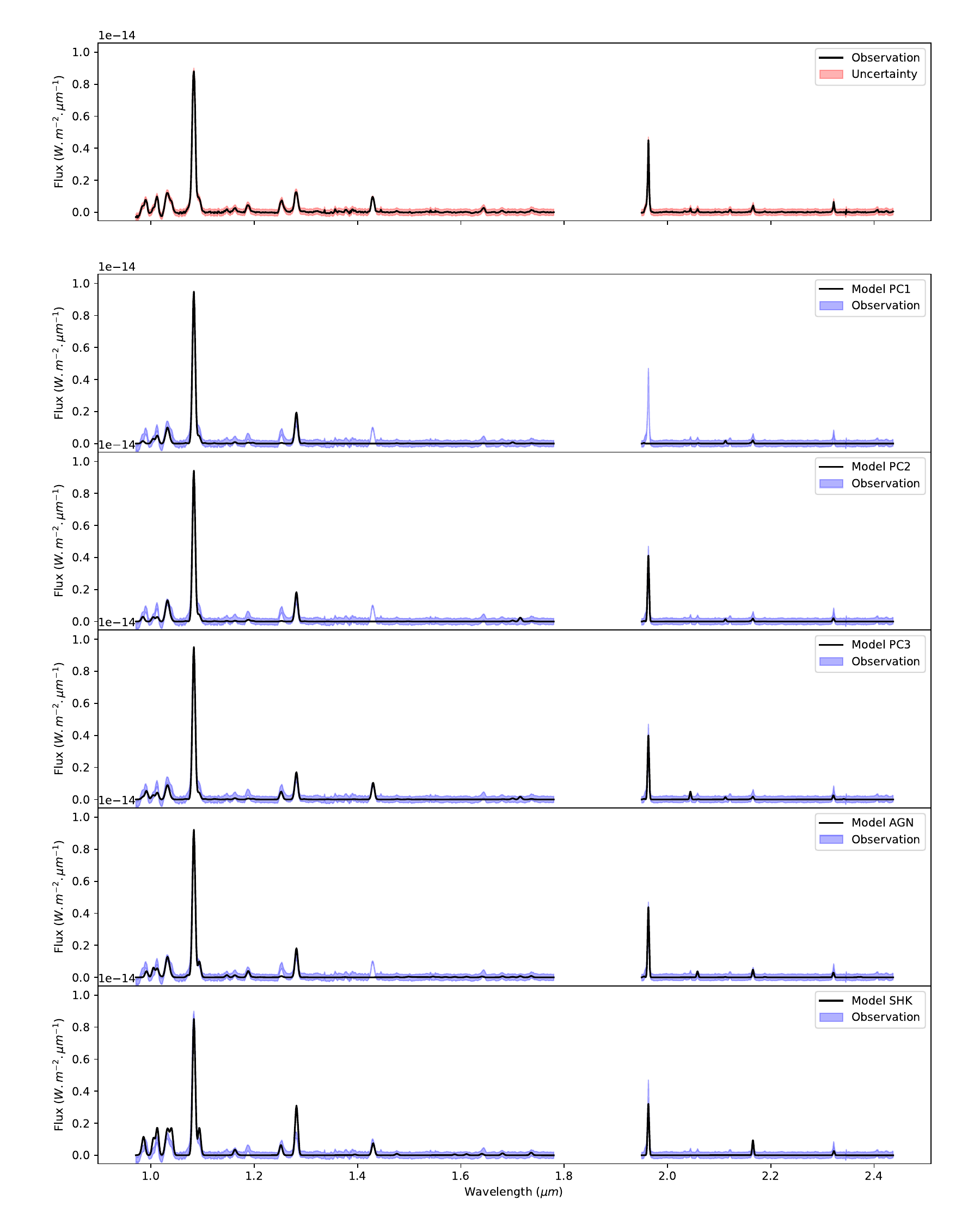}
  \caption{Cloud D: Comparison between the observed spectrum and the various models. The top panel shows the observed spectrum in black with the estimated uncertainty overlaid in transparent red. The five following panels show the prediction from models PC1, PC2, PC3, AGN, and SHK, in that order, with the observation overlaid in transparent blue.}
  \label{cloud2_comps}
\end{figure*}

\begin{figure}[!b]
\centering
\includegraphics[width=\linewidth]{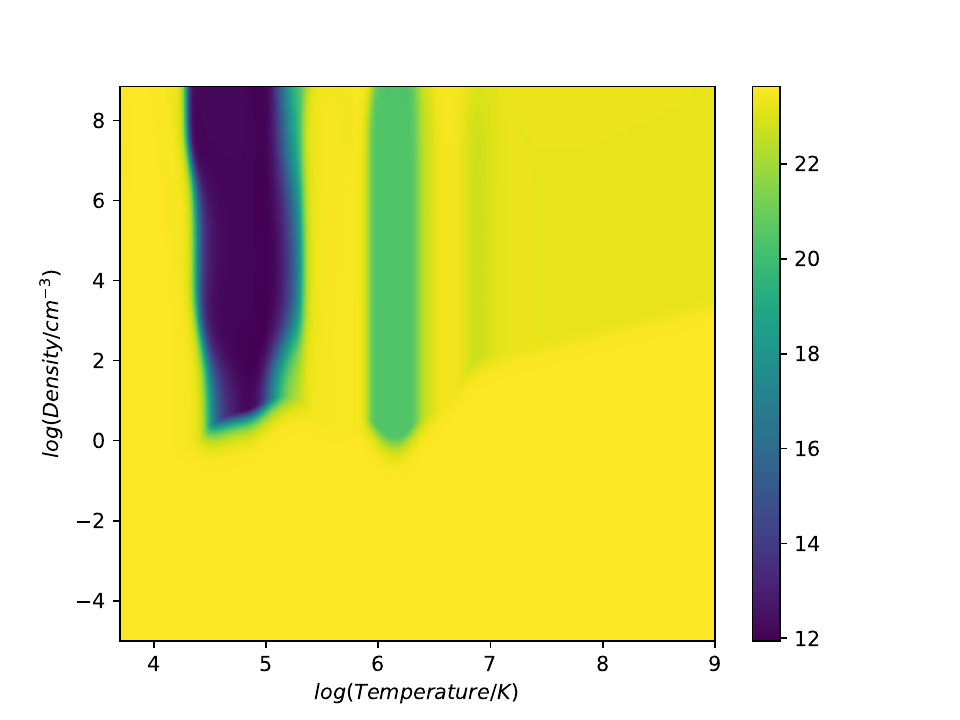}
\includegraphics[width=\linewidth]{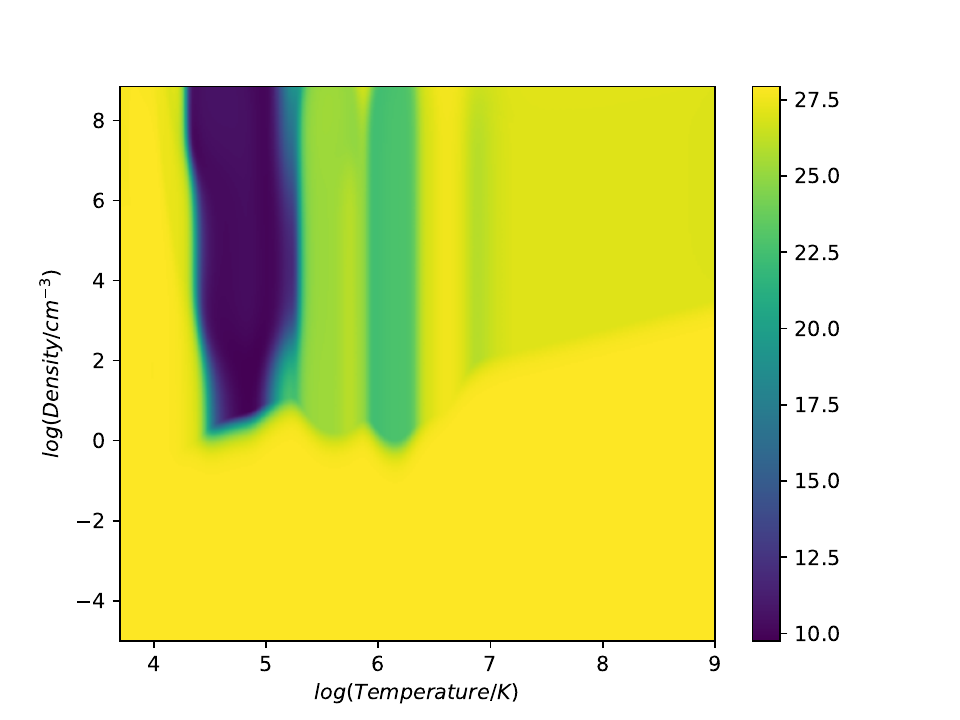}
\includegraphics[width=\linewidth]{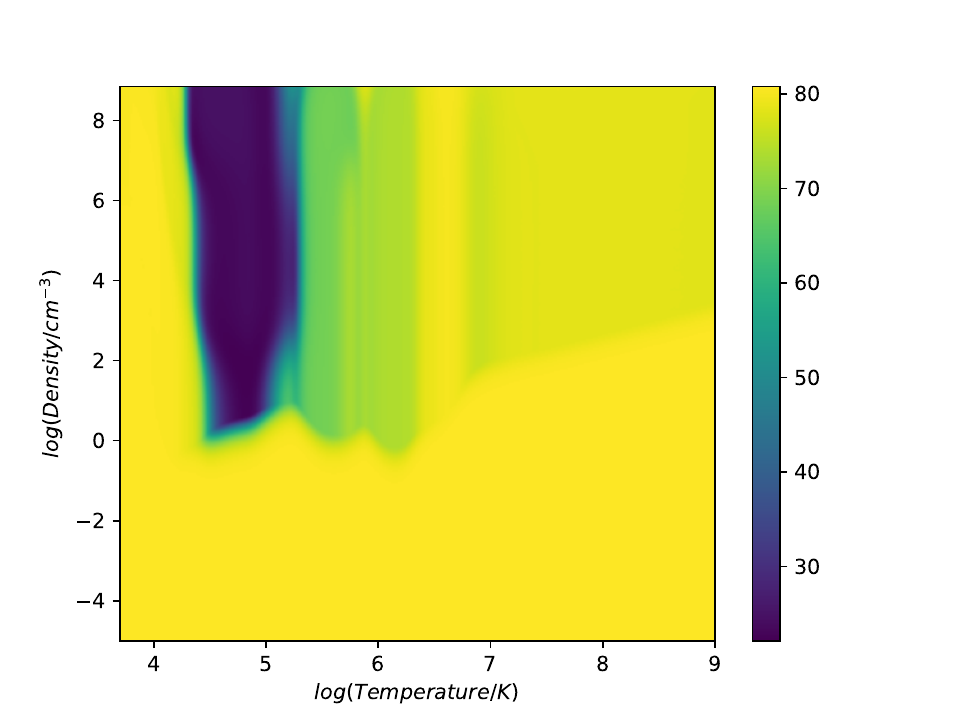}
  \caption{Model PC1: Reduced chi-square between the model and the observation spectra as a function of temperature and density for clouds B (top), C (middle), and D (bottom).}
     \label{k2_density_temp}
\end{figure}

\begin{figure}[!b]
\centering
\includegraphics[width=\linewidth]{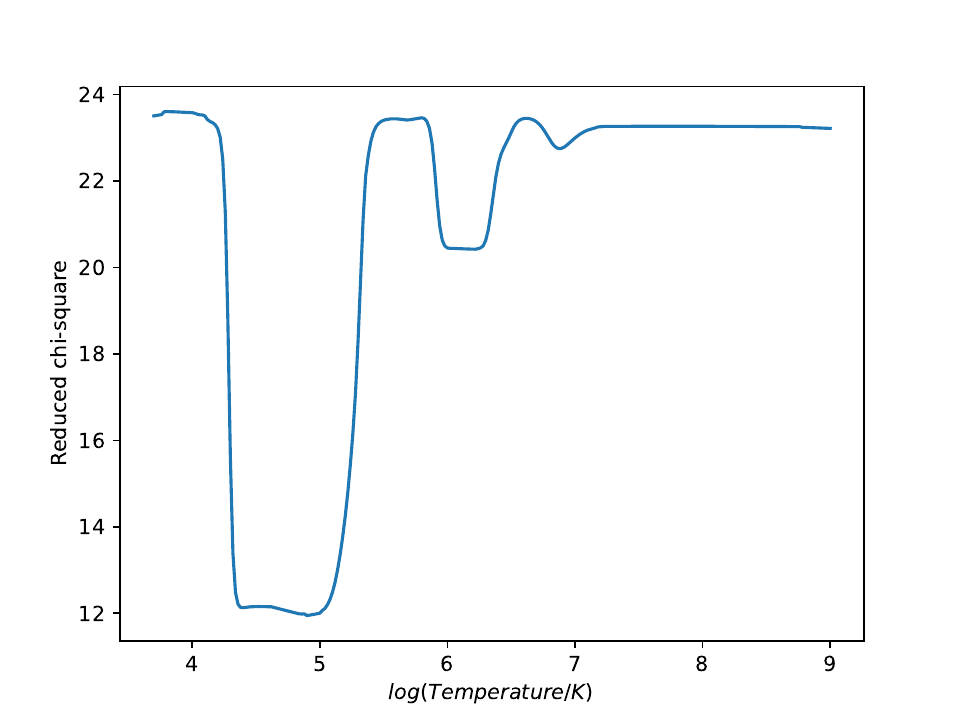}
\includegraphics[width=\linewidth]{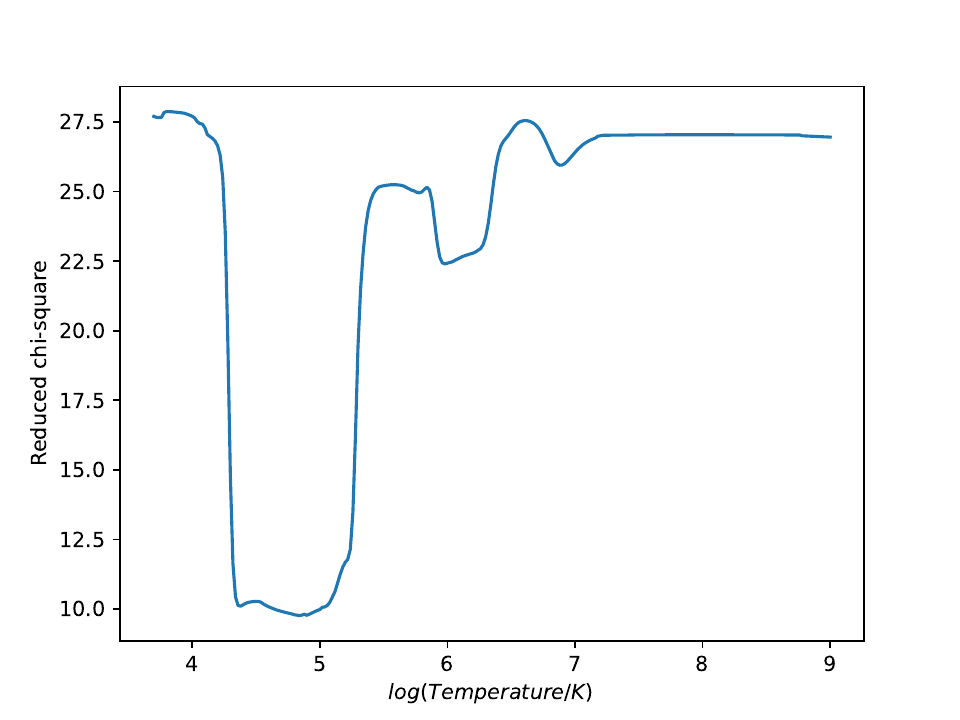}
\includegraphics[width=\linewidth]{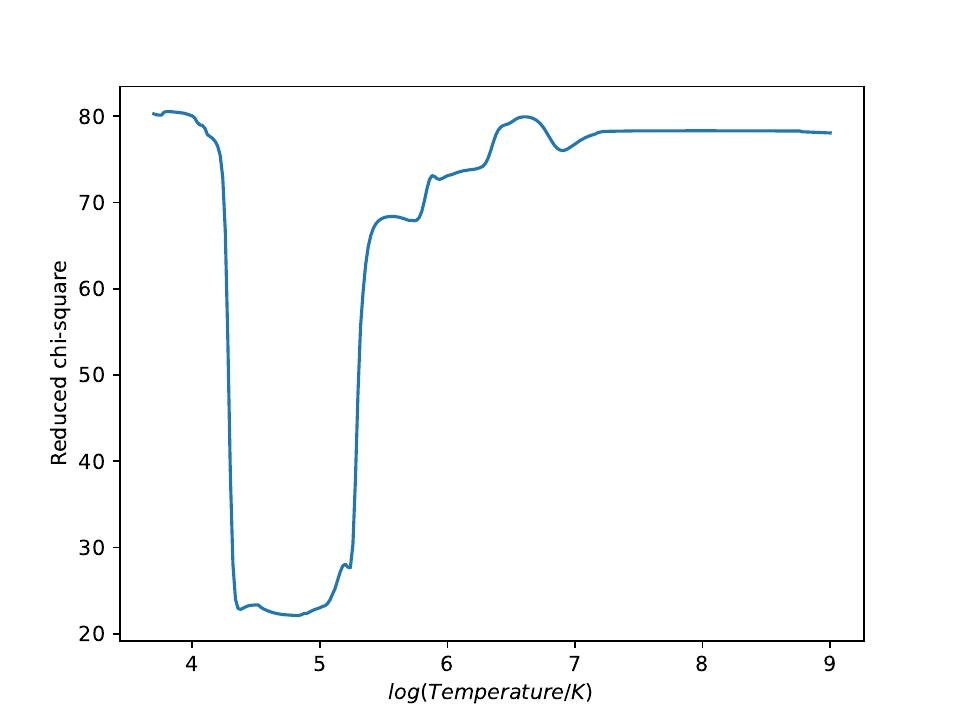}
  \caption{Model PC1: Reduced chi-square between the model and the observation spectra as a function of temperature for clouds B (top), C (middle), and D (bottom).}
  \label{k2_temp}
\end{figure}

\begin{figure}[!b]
\centering
\includegraphics[width=\linewidth]{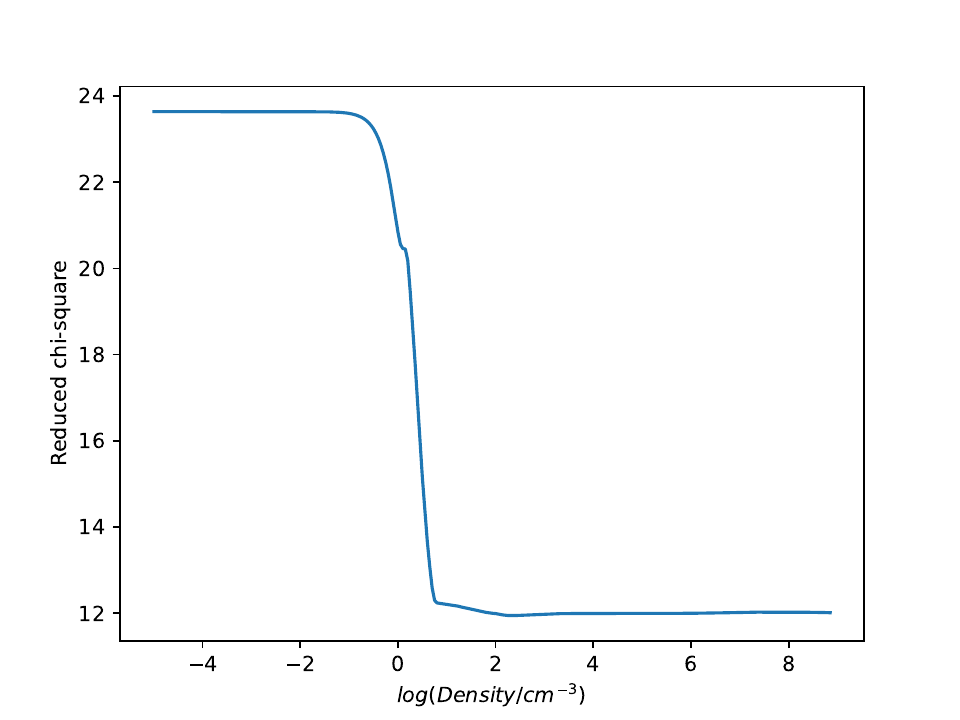}
\includegraphics[width=\linewidth]{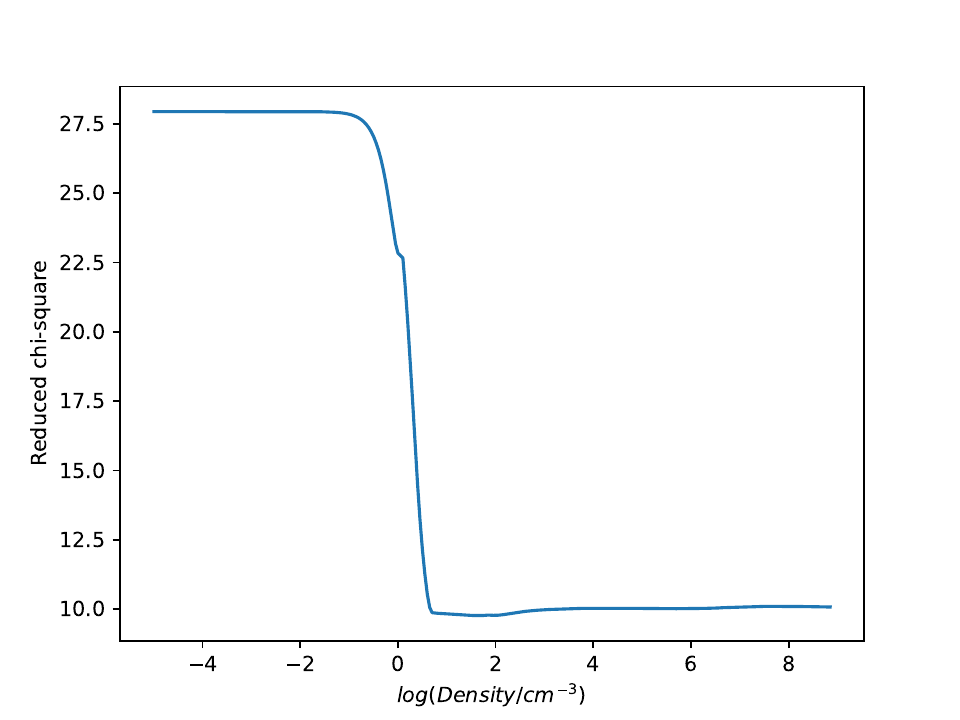}
\includegraphics[width=\linewidth]{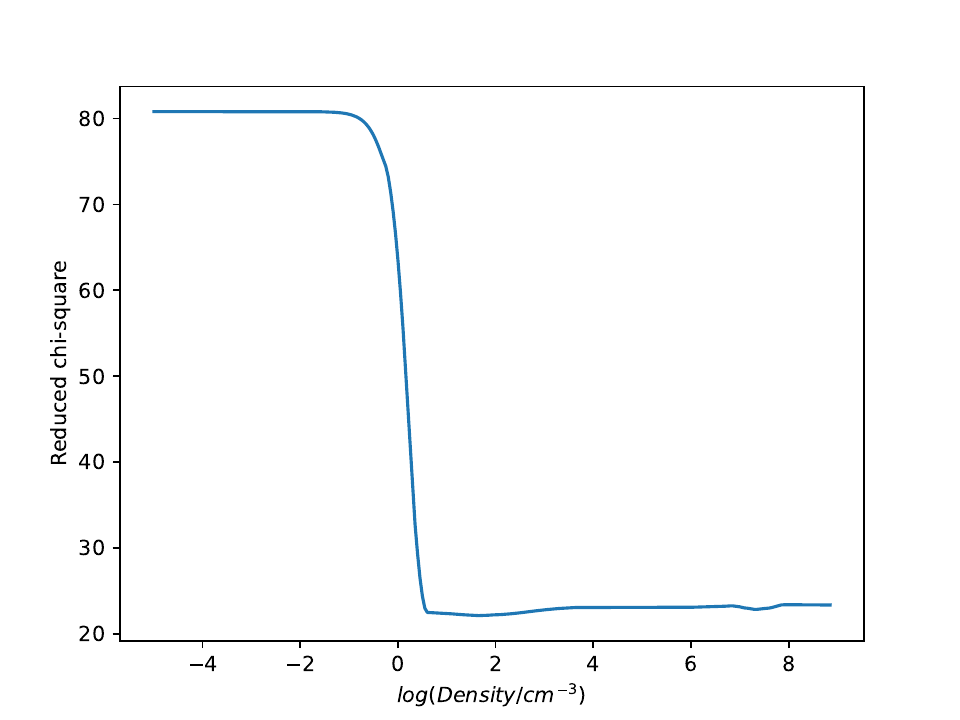}
  \caption{Model PC1: Reduced chi-square between the model and the observation spectra as a function of density for clouds B (top), C (middle), and D (bottom).}
  \label{k2_density}
\end{figure}

\begin{figure}[!b]
\centering
\includegraphics[width=\linewidth]{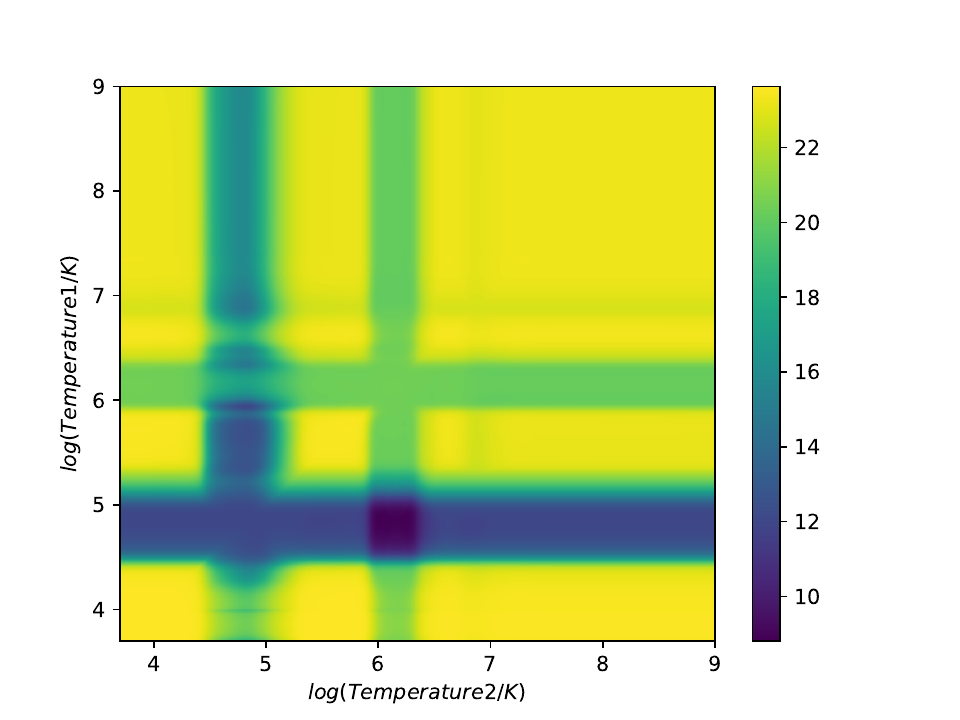}
\includegraphics[width=\linewidth]{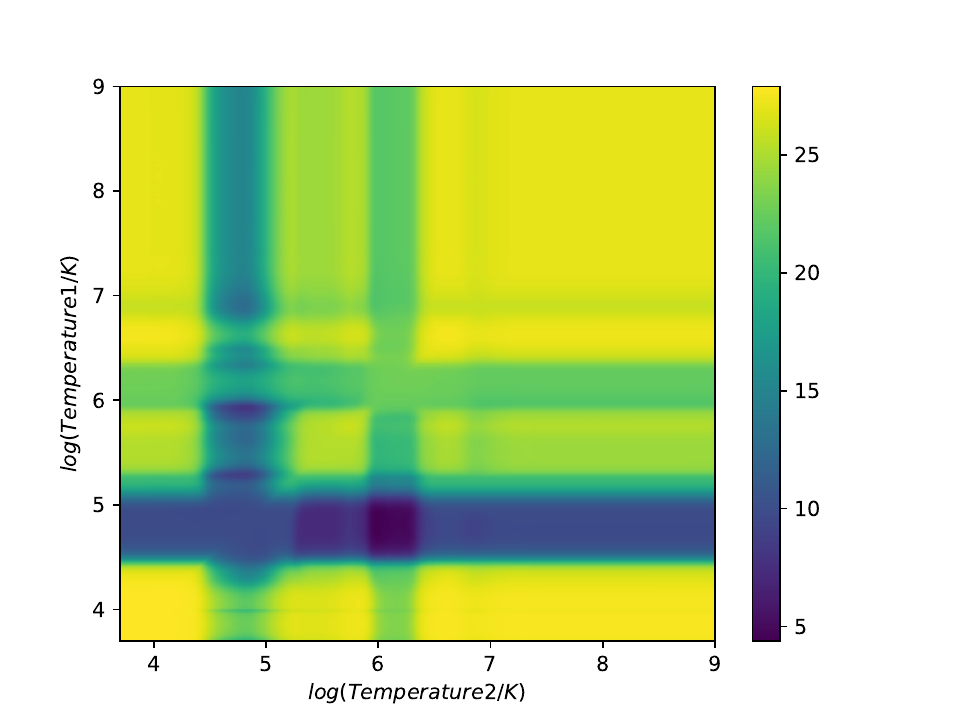}
\includegraphics[width=\linewidth]{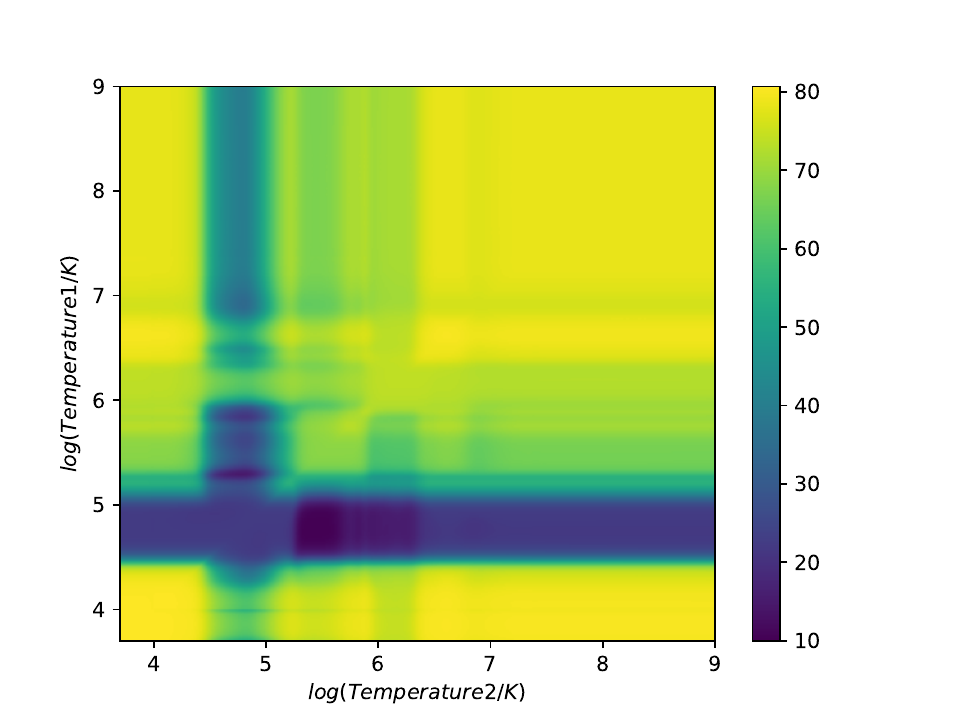}
  \caption{Model PC2: Reduced chi-square between the model and the observation spectra as a function of $T_1$ and $T_2$ for clouds B (top), C (middle), and D (bottom).}
     \label{k2_temp12}
\end{figure}

\begin{figure}[!b]
\centering
\includegraphics[width=\linewidth]{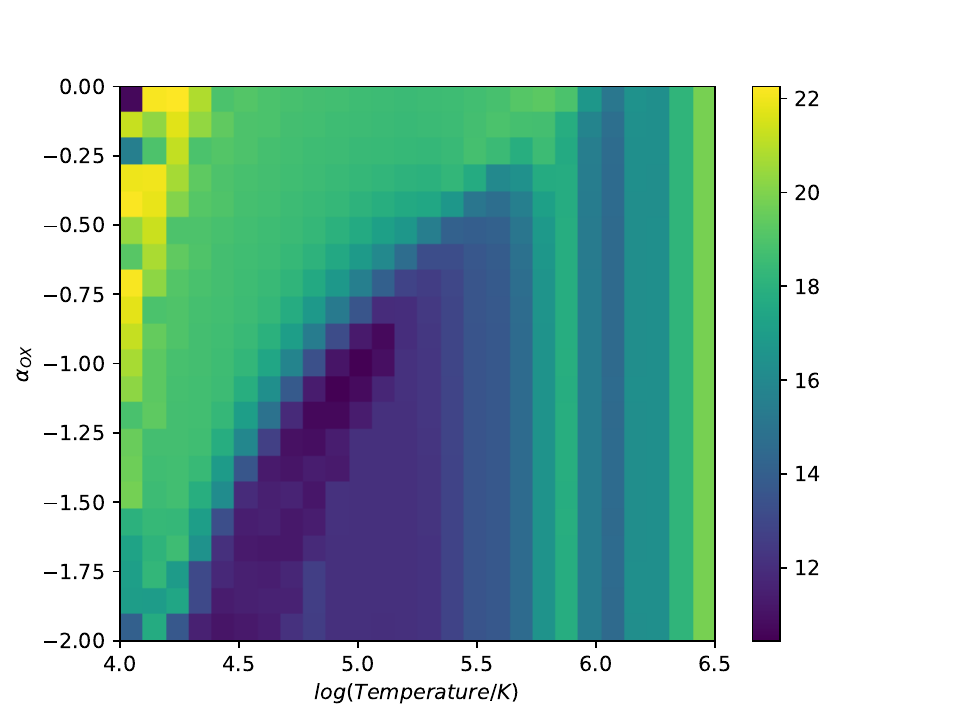}
\includegraphics[width=\linewidth]{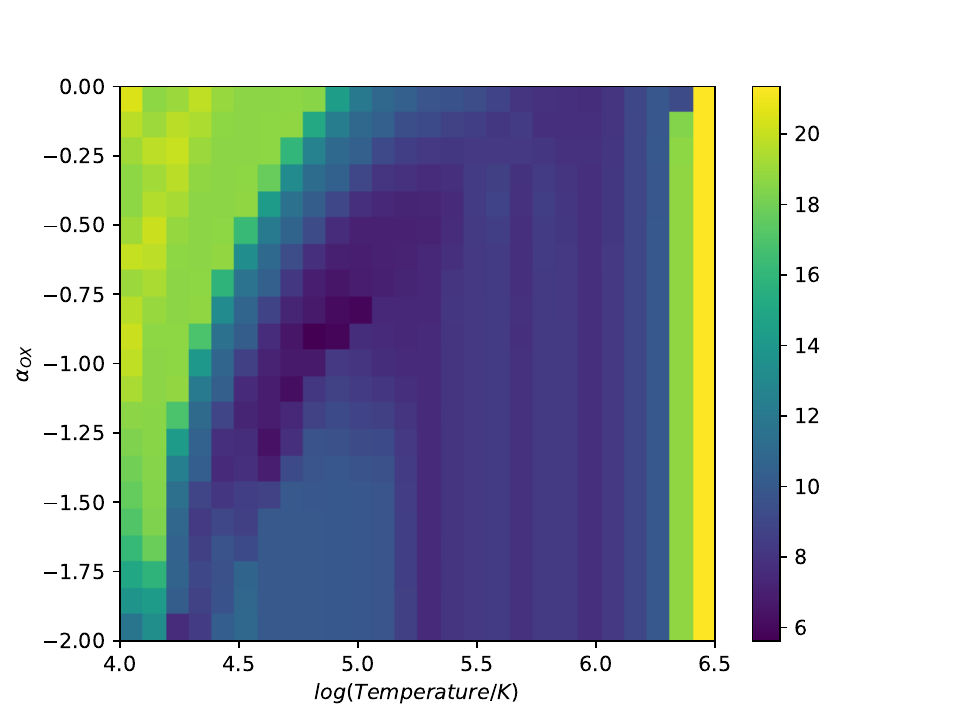}
\includegraphics[width=\linewidth]{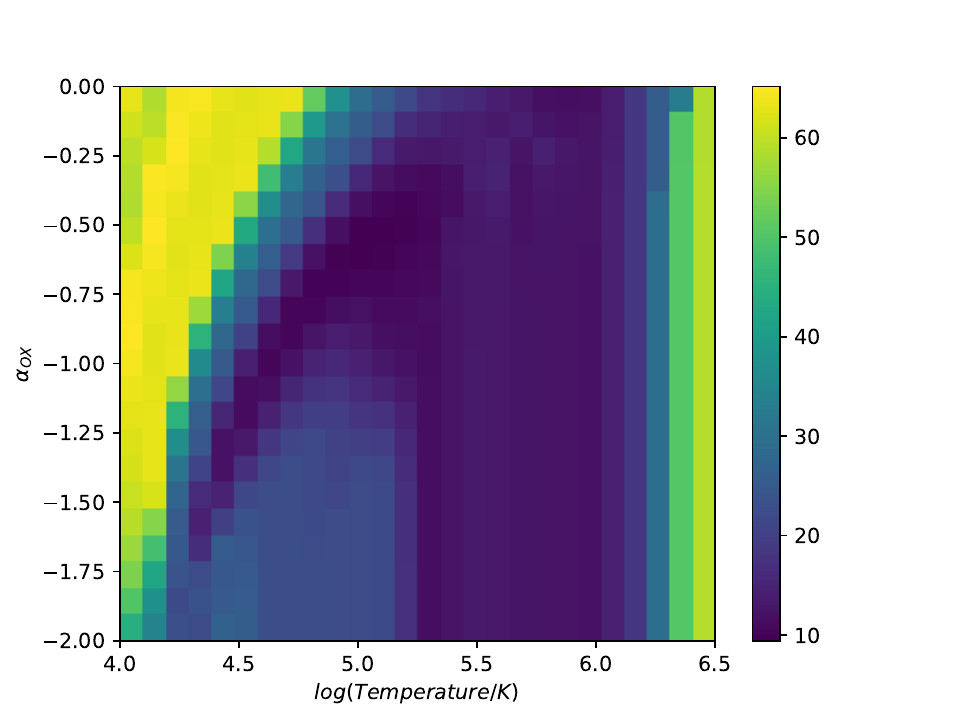}
  \caption{Model AGN: Reduced chi-square between the model and the observation spectra as a function of temperature and $\alpha_{OX}$ for clouds B (top), C (middle), and D (bottom).}
     \label{AGN_k2_temp_dens}
\end{figure}

\begin{figure}[!b]
\centering
\includegraphics[width=\linewidth]{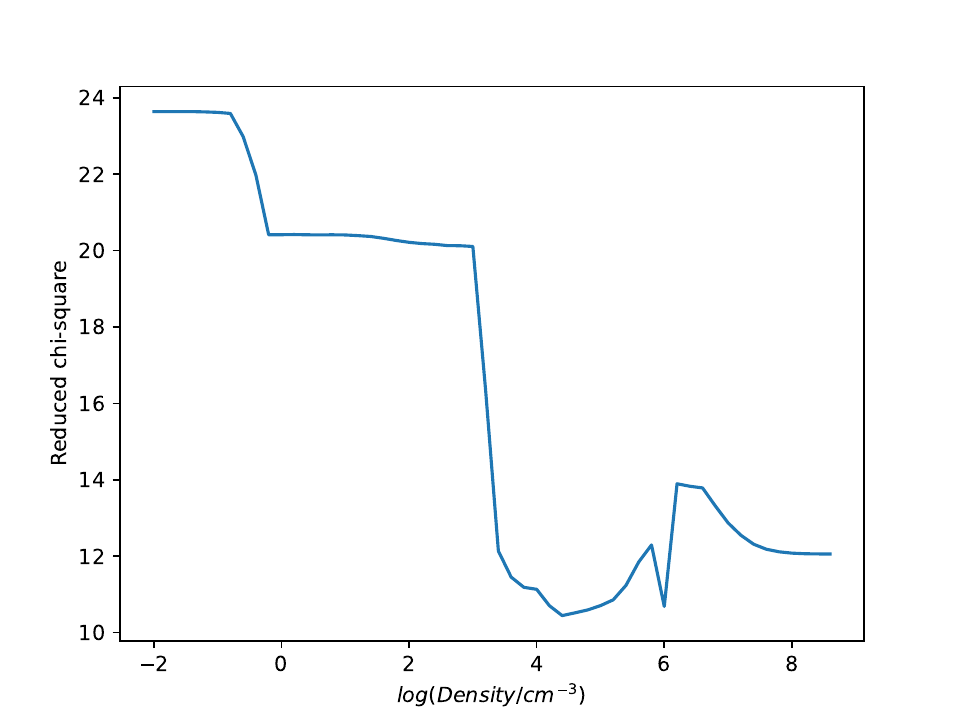}
\includegraphics[width=\linewidth]{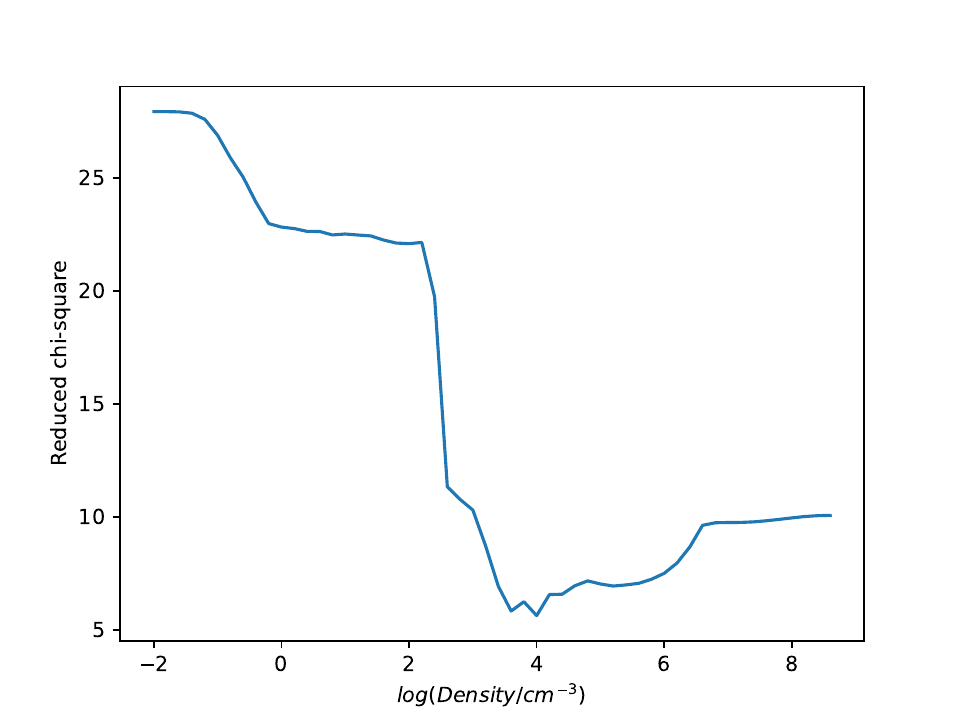}
\includegraphics[width=\linewidth]{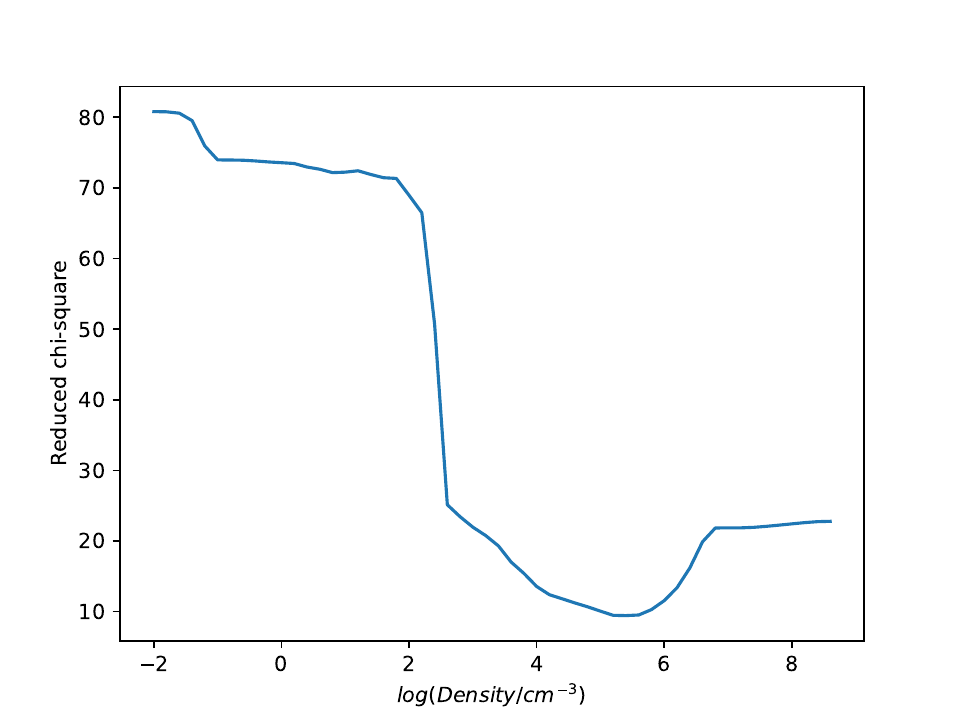}
  \caption{Model AGN: Reduced chi-square between the model and the observation spectra as a function of density for clouds B (top), C (middle), and D (bottom).}
     \label{AGN_k2_dens}
\end{figure}

\begin{figure}[!b]
\centering
\includegraphics[width=\linewidth]{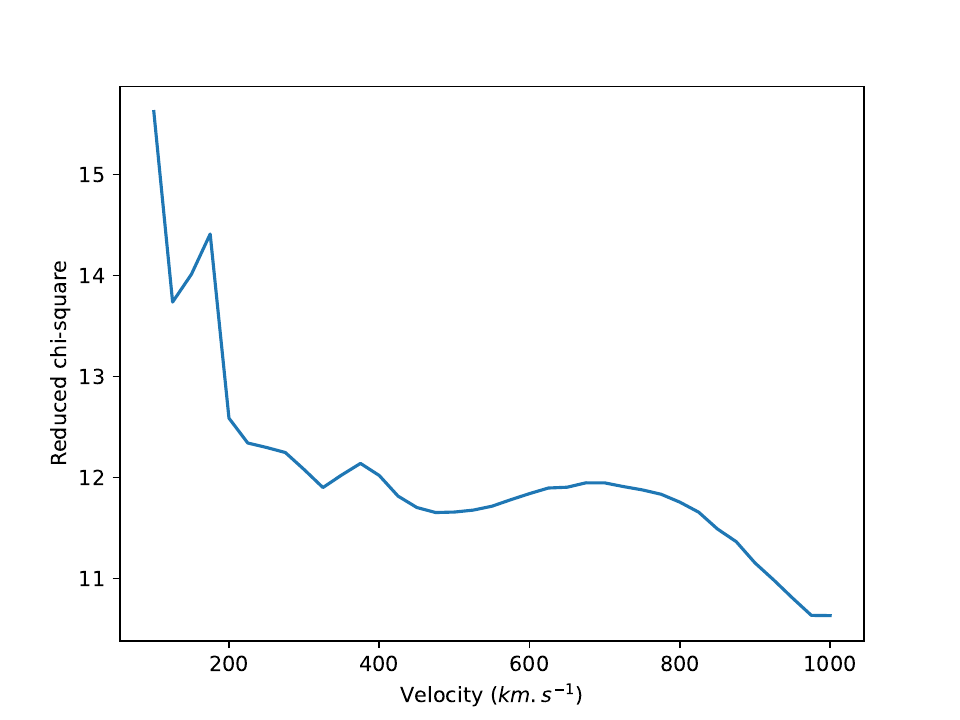}
\includegraphics[width=\linewidth]{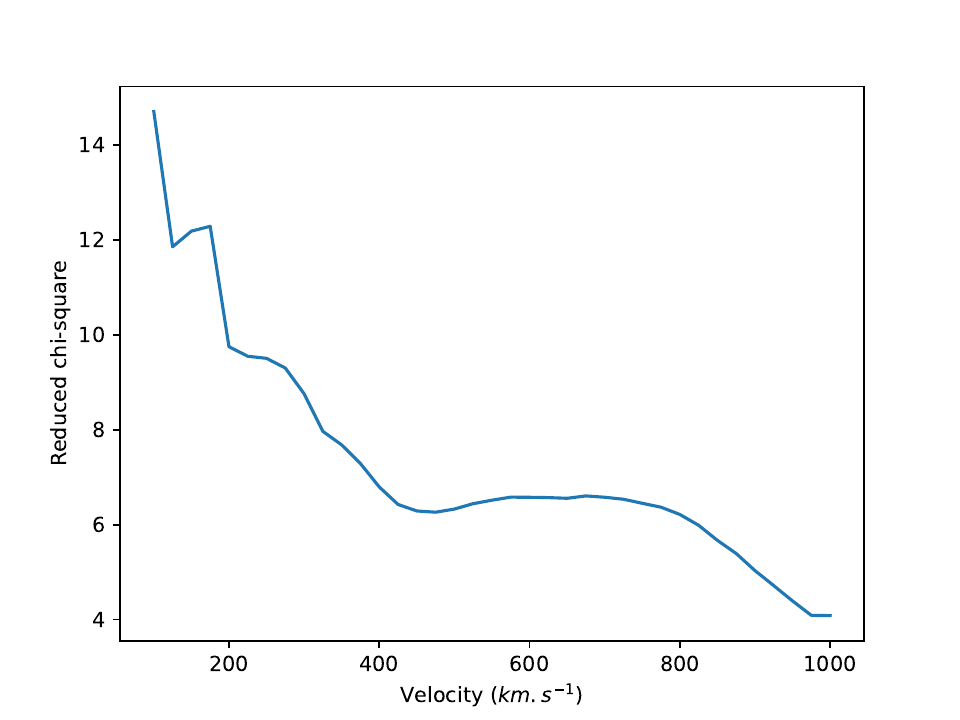}
\includegraphics[width=\linewidth]{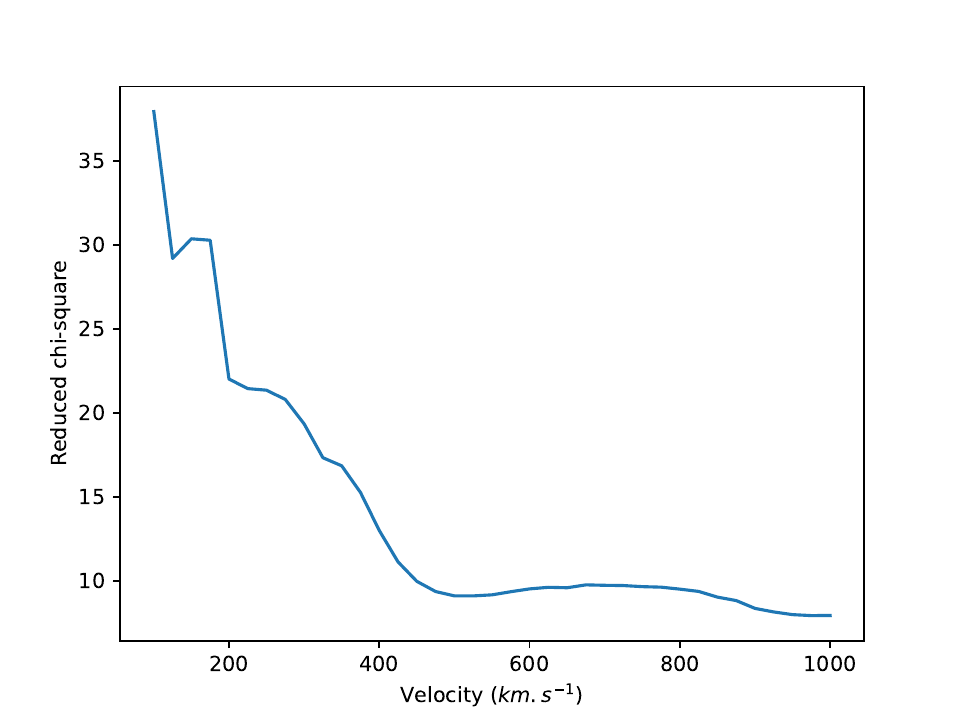}
  \caption{Model SHK: Reduced chi-square between the model and the observation spectra as a function of the shock velocity for clouds B (top), C (middle), and D (bottom).}
     \label{shock_k2_velocity}
\end{figure}

\begin{figure}[!b]
\centering
\includegraphics[width=\linewidth]{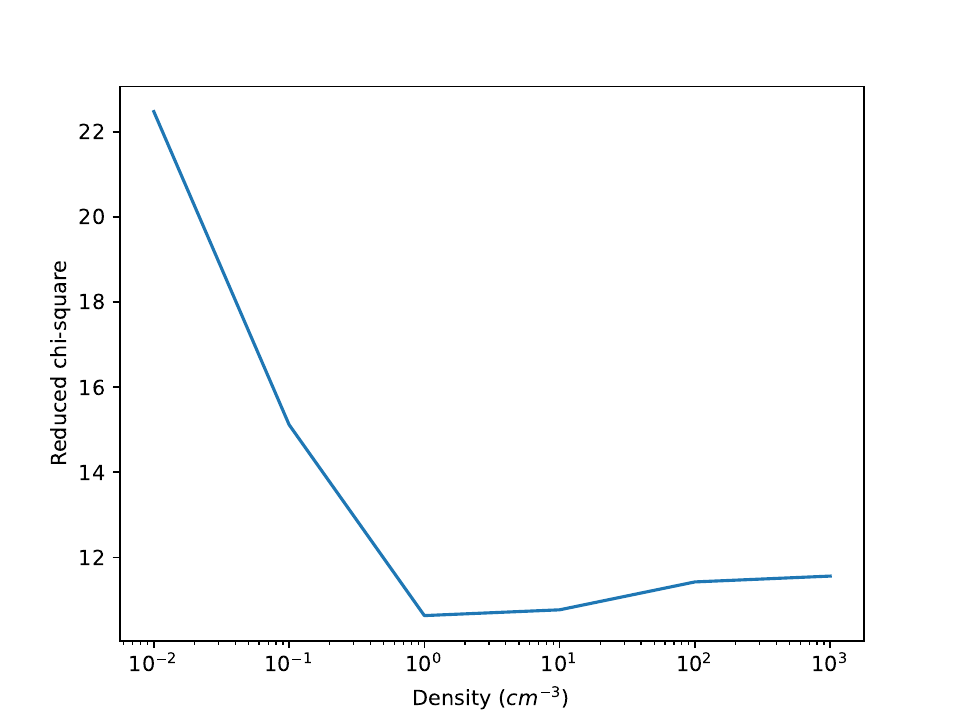}
\includegraphics[width=\linewidth]{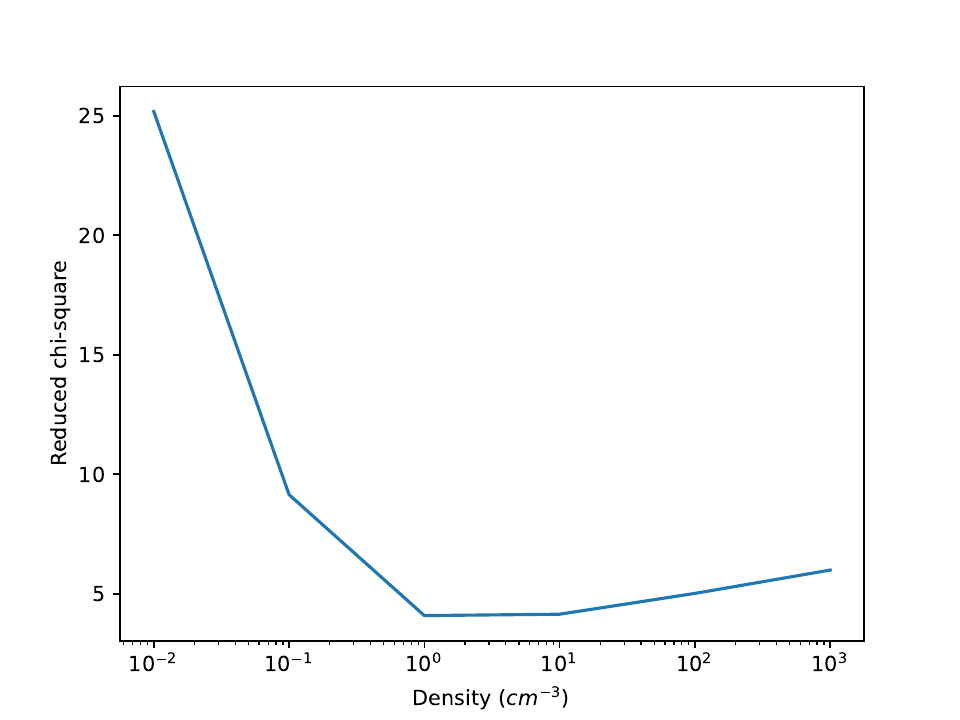}
\includegraphics[width=\linewidth]{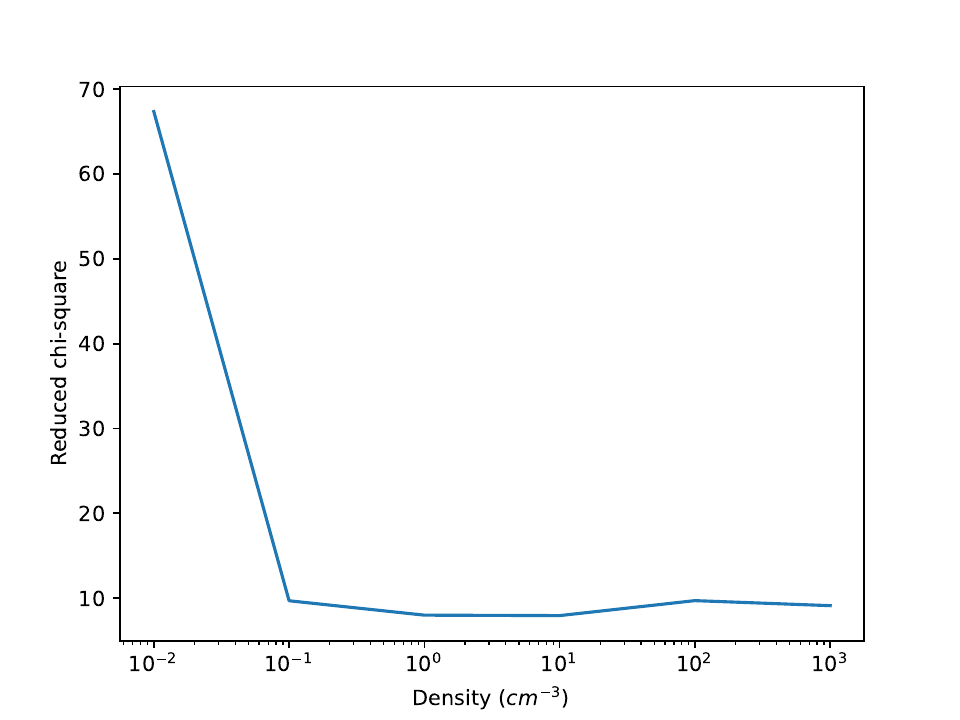}
  \caption{Model SHK: Reduced chi-square between the model and the observation spectra as a function of the preshock density for clouds B (top), C (middle), and D (bottom).}
     \label{shock_k2_density}
\end{figure}

\begin{figure}[!b]
\centering
\includegraphics[width=\linewidth]{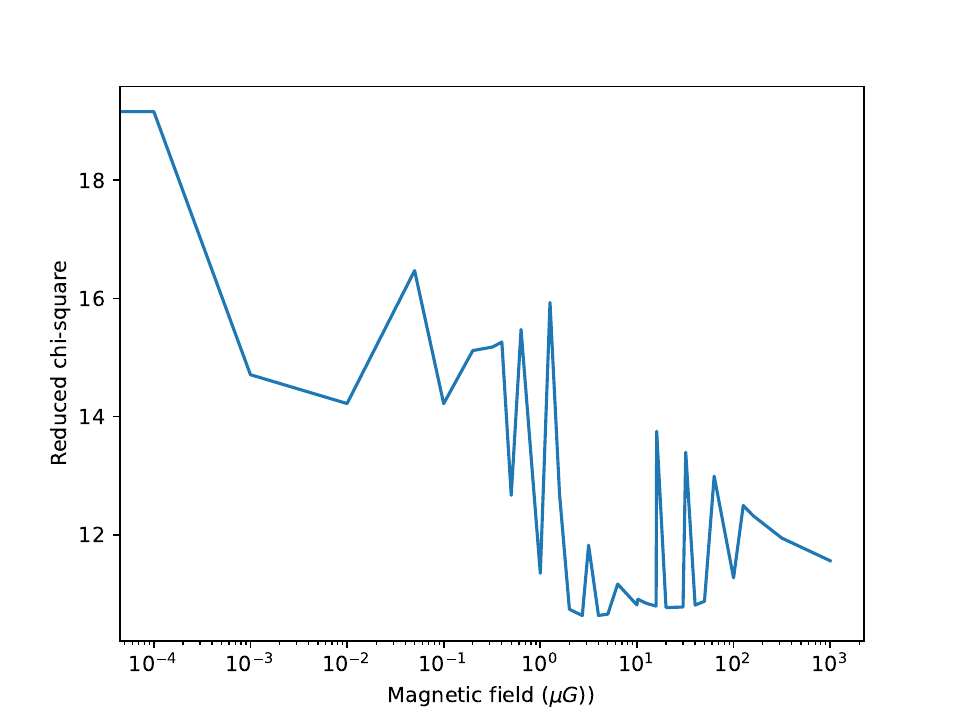}
\includegraphics[width=\linewidth]{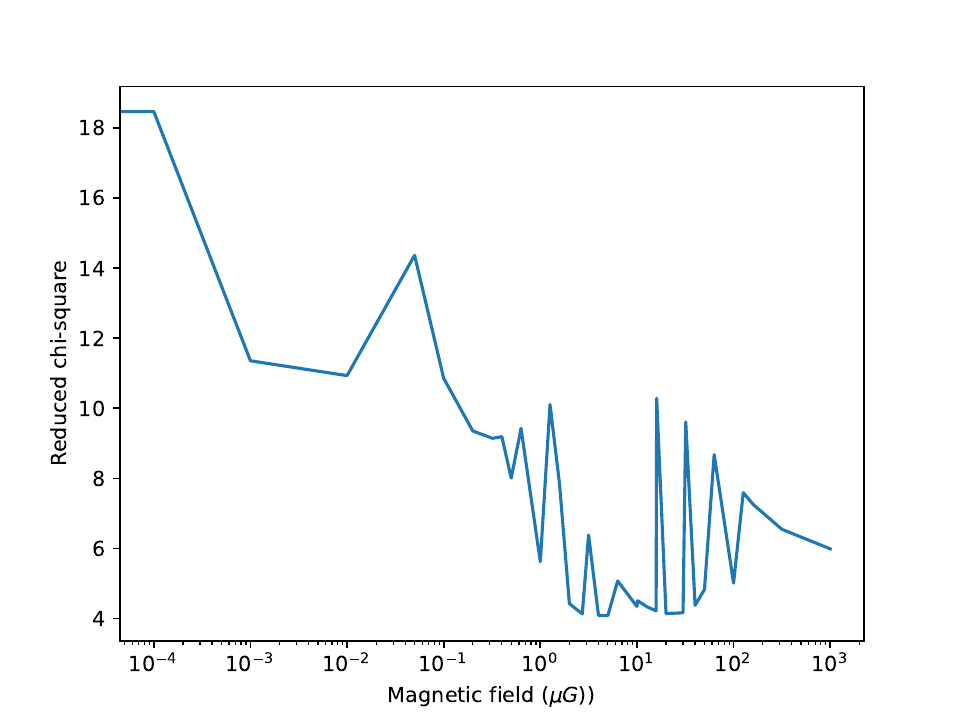}
\includegraphics[width=\linewidth]{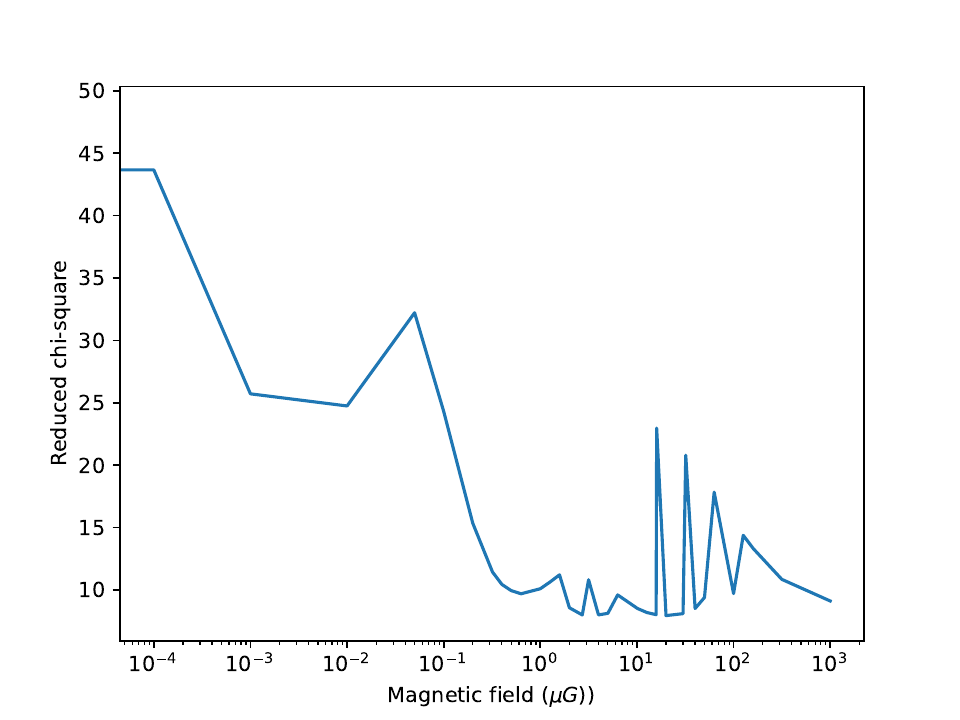}
  \caption{Model SHK: Reduced chi-square between the model and the observation spectra as a function of the preshock transverse magnetic field for clouds B (top), C (middle), and D (bottom).}
     \label{shock_k2_mag}
\end{figure}

The following appendices contain additional information supporting the conclusions drawn in the main body of the
text.
\bigbreak
For each ionized region, Appendix \ref{app:comps} presents a comparison between the observed spectrum and the various models discussed in the analysis: PC1, PC2, PC3, AGN, and SHK. Figure \ref{cloud0_comps} presents the results for cloud B, while Figs. \ref{cloud1_comps} and \ref{cloud2_comps} present the results for clouds C and D, respectively. Appendix \ref{app:PC12} presents the reduced chi-square maps for models PC1 and PC2. For model PC1, Fig. \ref{k2_density_temp} shows the 2D maps as a function of density and temperature for clouds B, C, and D, while Figs. \ref{k2_temp} and \ref{k2_density} show the 1D plots for the same parameters. For model PC2, Fig. \ref{k2_temp12} shows the 2D maps of the reduced chi-square as a function of $T_1$ and $T_2$ for the three regions. Appendix \ref{app:AGN} presents the reduced chi-square for the central photoionization model AGN. Figure \ref{AGN_k2_temp_dens} shows the 2D maps for the parameters of the central source ($\alpha_{OX}$ and $T$) for the three regions, while Fig. \ref{AGN_k2_dens} shows the chi-square as a function of the density of the cloud. Appendix \ref{app:SHK} presents the reduced chi-square for the fast radiative shock model SHK. Figures \ref{shock_k2_velocity}, \ref{shock_k2_density}, and \ref{shock_k2_mag} show the 1D chi-square as a function of shock velocity, preshock density, and transverse magnetic field, respectively, for all three clouds. In Appendix \ref{app:tables_SHK}, we compare the measured flux of each line in clouds B, C, and D with the prediction of the SHK model. Appendix \ref{app:temp_dens_lines} presents the emitting conditions (temperature, density) for a few selected emission lines of interest: [Fe II], [Si VI], [Si X], and [C I]. Finally, Appendix \ref{app:velocities} presents the [Si VI] Doppler shift maps for clouds C and D in Figs. \ref{cloud1_velocity} and \ref{cloud2_velocity}, respectively.

\section{Comparison between models and observations}

\label{app:comps}

\clearpage
\newpage
\mbox{~}

\section{Physical conditions}

\label{app:PC12}

\subsection{Single temperature and density}

\subsection{Double temperature, $n = 2\ cm^{-3}$}

\clearpage
\newpage
\mbox{~}

\section{Central photoionization}
\label{app:AGN}

\clearpage
\newpage
\mbox{~}
\section{Fast radiative shock}
\label{app:SHK}

\onecolumn
\section{Line fluxes from model SHK}
\label{app:tables_SHK}
\centering
\captionof{table}{\protect\label{table_comp_B} Cloud B: Comparison between the observation and the best SHK model.}
\begin{tabular}{c|c|c|c|c}
Line & Rest & Ionization & Measured & Model \\
Identification & Wavelength & Energy & Flux & Flux \\
  & (nm)  & (eV) & $(10^{-18}\ W.m^{-2}$) & $(10^{-18}\ W.m^{-2}$)\\
\hline
{[C I]} & 985.1 & 11.3 & 4.5 $\pm$ 1.4 & 20.9 \\
{[S VII]} & 986.9 & 281.0 & 7.8 $\pm$ 1.8 & 9.3 \\
{H I} & 1004.9 & 13.6 & 8.5 $\pm$ 2.0 & 16.0 \\
{He II} & 1012.4 & 54.4 & 11.4 $\pm$ 2.2 & 25.2 \\
{[S II]*} & 1032.0 & 23.3 & 10.3 $\pm$ 0.9 & 28.1 \\
{[N I]*} & 1039.8 & 14.5 & 5.3 $\pm$ 0.9 & 12.8 \\
{He I} & 1083.3 & 24.6 & 179.9 $\pm$ 7.6 & 130.1 \\
{H I} & 1093.8 & 13.6 & 7.1 $\pm$ 4.3 & 26.1 \\
{[P II]} & 1146.0 & 19.8 & 2.1 $\pm$ 0.5 & - \\
{He II} & 1167.3 & 54.4 & 3.2 $\pm$ 0.5 & 5.0 \\
{[P II]} & 1188.6 & 19.8 & 2.9 $\pm$ 0.5 & - \\
{[S IX]} & 1251.7 & 379.8 & 12.3 $\pm$ 0.6 & 10.6 \\
{H I} & 1281.8 & 13.6 & 24.8 $\pm$ 0.9 & 47.8 \\
{[Si X]} & 1430.2 & 401.4 & 24.1 $\pm$ 5.0 & 13.3 \\
\end{tabular}

\bigbreak

\centering
\captionof{table}{\protect\label{table_comp_C} Cloud C: Comparison between the observation and the best SHK model.}
\begin{tabular}{c|c|c|c|c}
Line & Rest & Ionization & Measured & Model \\
Identification & Wavelength & Energy & Flux & Flux \\
  & (nm)  & (eV) & $(10^{-18}\ W.m^{-2}$) & $(10^{-18}\ W.m^{-2}$)\\
\hline
{[C I]} & 985.1 & 11.3 & 0.7 $\pm$ 0.5 & 7.8 \\
{[S VII]} & 986.9 & 281.0 & 11.1 $\pm$ 0.8 & 7.3 \\
{H I} & 1004.9 & 13.6 & 3.0 $\pm$ 0.8 &  12.2\\
{He II} & 1012.4 & 54.4 & 9.5 $\pm$ 1.2 & 20.4 \\
{[S II]*} & 1032.0 & 23.3 & 13.1 $\pm$ 0.6 & 20.4 \\
{[N I]*} & 1039.8 & 14.5 & 7.8 $\pm$ 0.6 & 12.5 \\
{He I} & 1083.3 & 24.6 & 109.2 $\pm$ 2.6 & 92.5 \\
{H I} & 1093.8 & 13.6 & 13.2 $\pm$ 1.5 & 19.8 \\
{[P II]} & 1146.0 & 19.8 & 1.5 $\pm$ 0.2 & - \\
{He II} & 1167.3 & 54.4 & 4.0 $\pm$ 0.3 & 4.6 \\
{[P II]} & 1188.6 & 19.8 & 4.9 $\pm$ 0.3 & - \\
{[S IX]} & 1251.7 & 379.8 & 11.3 $\pm$ 0.5 & 8.2 \\
{H I} & 1281.8 & 13.6 & 16.3 $\pm$ 0.6 &  35.8\\
{[Si X]} & 1430.2 & 401.4 & 16.9 $\pm$ 0.5 & 10.3 \\
{[Fe II]} & 1643.6 & 16.2 & 2.5 $\pm$ 0.2 & / \\
{H I} & 1680.7 & 13.6 & 1.2 $\pm$ 0.2 & 1.5 \\
{H I} & 1736.2 & 13.6 & 1.8 $\pm$ 0.2 & 2.0 \\
{[Si VI]} & 1960.2 & 205.3 & 24.8 $\pm$ 1.7 & 19.8 \\
{He II} & 2033.0 & 24.6 & 1.1 $\pm$ 0.2 & 0.0 \\
{[Al IX]} & 2044.4 & 24.6 & 2.1 $\pm$ 0.2 & 0.0 \\
{He I} & 2058.1 & 24.6 & 0.4 $\pm$ 0.2 & 0.0 \\
{$H_2$} & 2121.2 & 0.0 & 0.7 $\pm$ 0.2 & - \\
{H I} & 2165.5 & 13.6 & 1.8 $\pm$ 0.3 & 6.1 \\
{[Ca VIII]} & 2322.2 & 147.3 & 4.3 $\pm$ 0.3 & 1.9 \\
\end{tabular}

\clearpage

\centering
\captionof{table}{\protect\label{table_comp_D} Cloud D: Comparison between the observation and the best SHK model.}
\begin{tabular}{c|c|c|c|c}
Line & Rest & Ionization & Measured & Model \\
Identification & Wavelength & Energy & Flux & Flux \\
  & (nm)  & (eV) & $(10^{-18}\ W.m^{-2}$) & $(10^{-18}\ W.m^{-2}$)\\
\hline
{[C I]} & 985.1 & 11.3 & 2.0 $\pm$ 0.3 & 4.8 \\
{[S VII]} & 986.9 & 281.0 & 5.1 $\pm$ 0.4 & 4.2 \\
{H I} & 1004.9 & 13.6 & 1.9 $\pm$ 0.5 & 7.1 \\
{He II} & 1012.4 & 54.4 & 5.5 $\pm$ 0.8 & 11.7 \\
{[S II]*} & 1032.0 & 23.3 & 9.3 $\pm$ 0.4 & 15.9 \\
{[N I]*} & 1039.8 & 14.5 & 4.6 $\pm$ 0.4 & 10.1 \\
{He I} & 1083.3 & 24.6 & 66.3 $\pm$ 0.9 & 58.1 \\
{H I} & 1093.8 & 13.6 & 6.1 $\pm$ 0.5 & 11.5 \\
{[P II]} & 1146.0 & 19.8 & 1.3 $\pm$ 0.1 & - \\
{He II} & 1167.3 & 54.4 & 2.1 $\pm$ 0.2 & 2.6 \\
{[P II]} & 1188.6 & 19.8 & 3.6 $\pm$ 0.2 & - \\
{[S IX]} & 1251.7 & 379.8 & 5.8 $\pm$ 0.2 & 4.4 \\
{H I} & 1281.8 & 13.6 & 10.4 $\pm$ 0.2 & 20.1 \\
{[Si X]} & 1430.2 & 401.4 & 6.6 $\pm$ 0.2 & 5.2 \\
{[Fe II]} & 1643.6 & 16.2 & 2.0 $\pm$ 0.1 & / \\
{H I} & 1680.7 & 13.6 & 0.7 $\pm$ 0.1 &  0.9\\
{H I} & 1736.2 & 13.6 & 1.1 $\pm$ 0.1 &  1.1\\
{[Si VI]} & 1960.2 & 205.3 & 15.1 $\pm$ 0.4 & 12.0 \\
{He II} & 2033.0 & 24.6 & 0.5 $\pm$ 0.1 & 0.0 \\
{[Al IX]} & 2044.4 & 24.6 & 0.9 $\pm$ 0.1 & 0.0 \\
{He I} & 2058.1 & 24.6 & 0.9 $\pm$ 0.1 & 0.0 \\
{$H_2$} & 2121.2 & 0.0 & 0.7 $\pm$ 0.0 & - \\
{H I} & 2165.5 & 13.6 & 1.9 $\pm$ 0.1 & 3.5 \\
{[Ca VIII]} & 2322.2 & 147.3 & 1.8 $\pm$ 0.1 & 1.0 \\
\end{tabular}

\twocolumn

\clearpage
\section{Physical conditions for selected emission lines}
\label{app:temp_dens_lines}
\begin{figure}
\centering
\includegraphics[width=\linewidth]{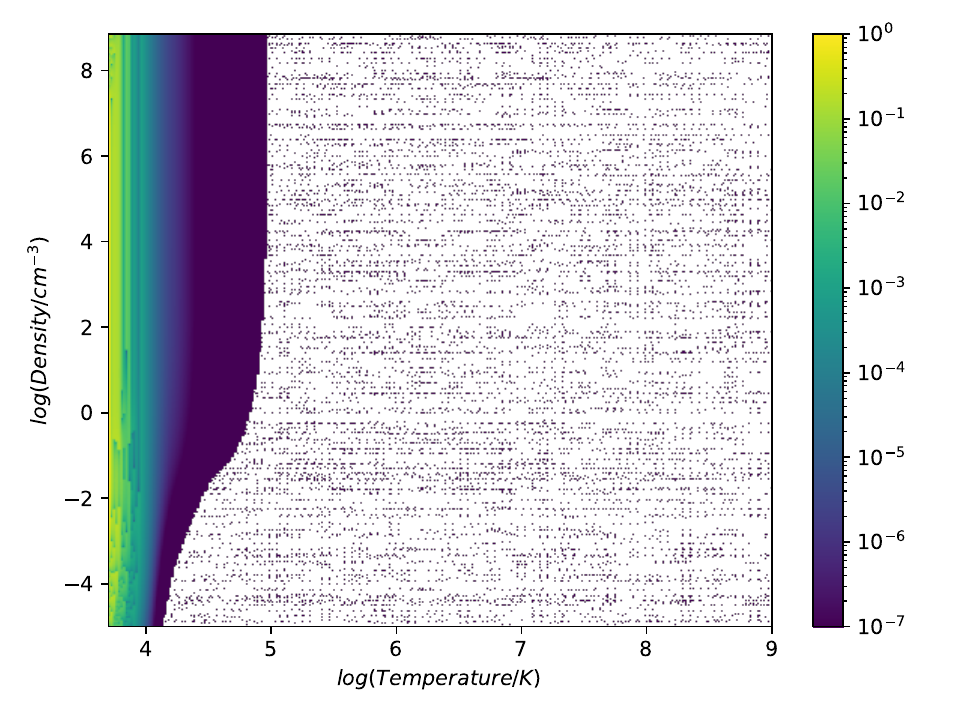}
  \caption{Emitting conditions for [Fe II] at 1.644 $\mu m$: $\frac{[Fe\ II]}{H\beta}$ normalized to one for its maximum value.}
  \label{feII_temp_dens}
\end{figure}

\begin{figure}
\centering
\includegraphics[width=\linewidth]{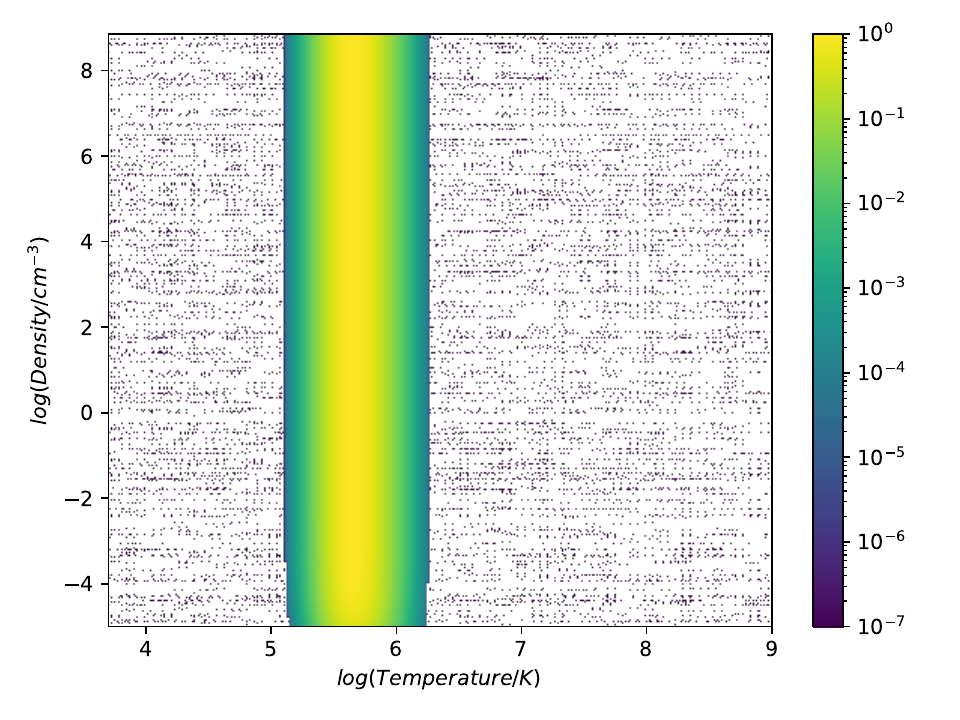}
  \caption{Emitting conditions for [Si VI] at 1.965 $\mu m$:  $\frac{[Si\ VI]}{H\beta}$ normalized to one for its maximum value.}
  \label{SiVI_temp_dens}
\end{figure}

\begin{figure}
\centering
\includegraphics[width=\linewidth]{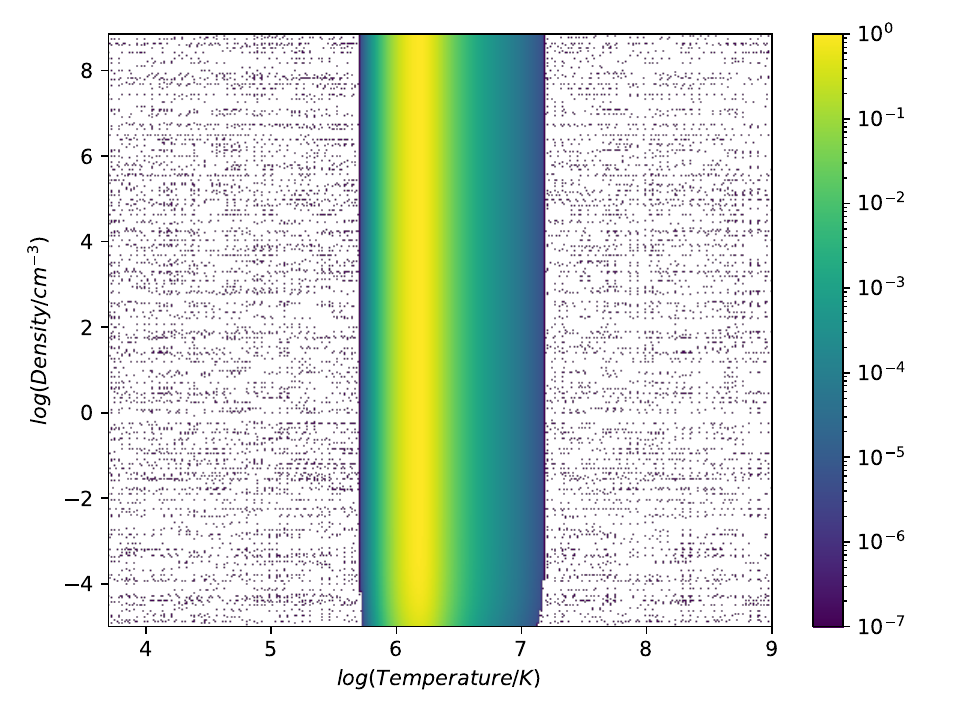}
  \caption{Emitting conditions for [Si X] at 1.430 $\mu m$:  $\frac{[Si\ X]}{H\beta}$ normalized to one for its maximum value.}
  \label{SiX_temp_dens}
\end{figure}

\begin{figure}
\centering
\includegraphics[width=\linewidth]{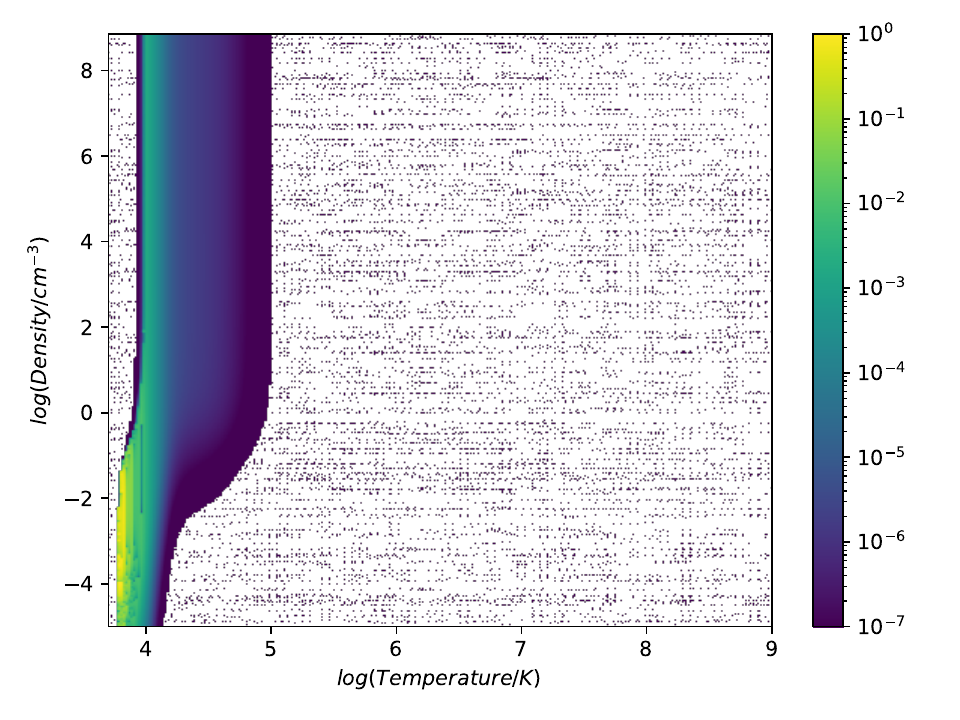}
  \caption{Emitting conditions for [C I] at 0.985 $\mu m$: $\frac{[C\ I]}{H\beta}$ normalized to one for its maximum value.}
  \label{CI_temp_dens}
\end{figure}

\clearpage
\newpage
\mbox{~}
\section{[Si VI] Doppler shift}
\label{app:velocities}
\begin{figure}
\centering
\includegraphics[width=\linewidth]{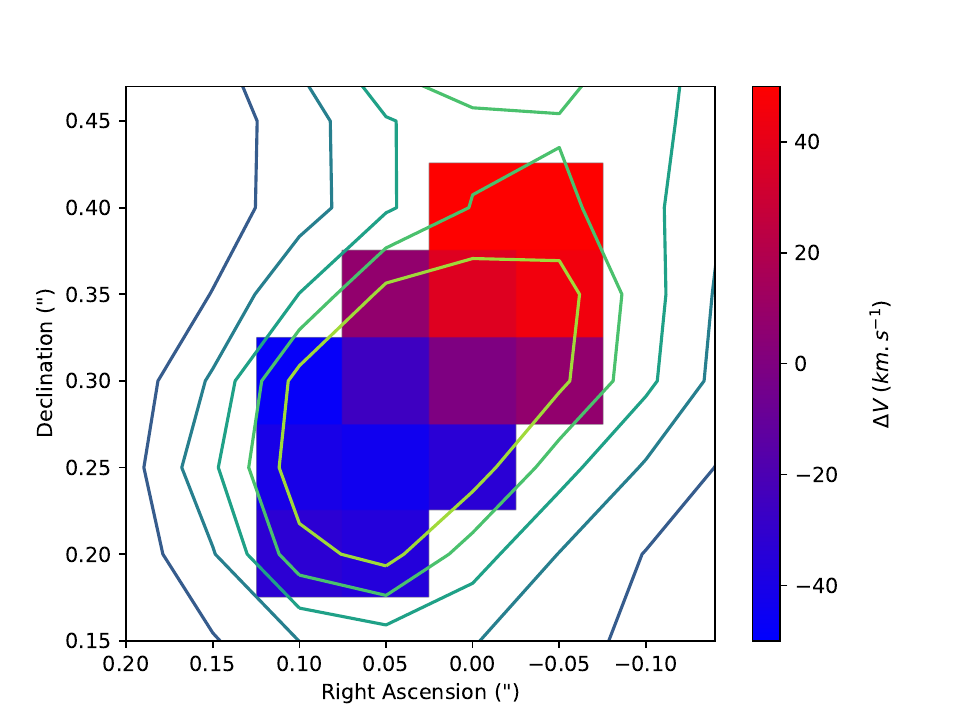}
  \caption{Cloud C: Velocity map based on the measured Doppler shift of the [Si VI] emission line.}
  \label{cloud1_velocity}
\end{figure}

\begin{figure}
\centering
\includegraphics[width=\linewidth]{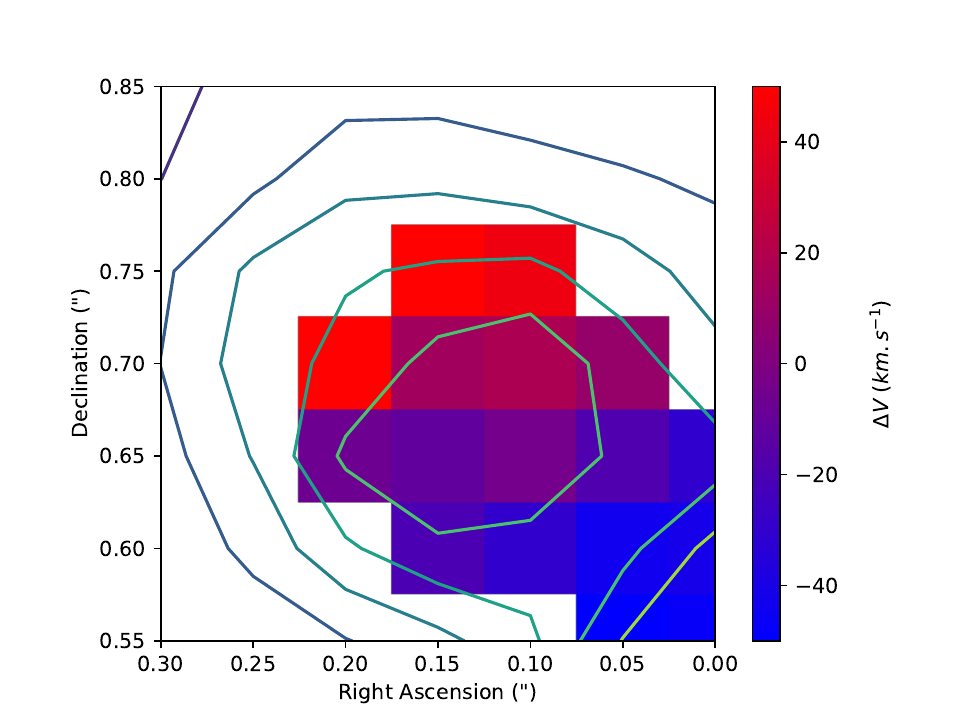}
  \caption{Cloud D: Velocity map based on the measured Doppler shift of the [Si VI] emission line.}
  \label{cloud2_velocity}
\end{figure}



\end{appendix}
\end{document}